\documentclass{aa}  
\pdfoutput=1
\usepackage{graphicx}

\usepackage{txfonts}

\usepackage{natbib}
\bibpunct{(}{)}{;}{a}{}{,} 

\usepackage[colorlinks=true, citecolor=blue, linkcolor=blue, urlcolor=blue]{hyperref}
\begin{document}

   \title{A new set of atmosphere and evolution models for cool T--Y brown dwarfs and giant exoplanets}


   \author{M. W. Phillips\inst{\ref{exeter}}\and
          P. Tremblin\inst{\ref{CEA}}\and
          I. Baraffe\inst{\ref{exeter}\and\ref{Lyon}}\and
          G. Chabrier\inst{\ref{exeter}\and\ref{Lyon}}\and
          N. F. Allard\inst{\ref{GEPI}\and\ref{IAP}}\and 
          F. Spiegelman\inst{\ref{Toulouse}}\and 
          J. M. Goyal\inst{\ref{exeter}\and\ref{Cornell}}\and
          B. Drummond\inst{\ref{exeter},\ref{Met}}\and
          E. H\'ebrard\inst{\ref{exeter}}}

   \institute{Astrophysics Group, University of Exeter, EX4 4QL, Exeter, UK \\ \email{mp537@exeter.ac.uk}\label{exeter}
   \and
   Maison de la Simulation, CEA, CNRS, Univ. Paris-Sud, UVSQ, Universit\'e Paris-Saclay, 91191 Gif-sur-Yvette, France \label{CEA}
   \and 
   Ecole Normale Sup\'erieure de Lyon, CRAL, UMR CNRS 5574, 69364 Lyon Cedex 07, France \label{Lyon}\and 
   GEPI, Observatoire de Paris PSL Research University, UMR 8111, CNRS, Sorbonne Paris Cit\'e,61, Avenue de l'Observatoire, F-75014 Paris, France \label{GEPI}, \and 
   Institut d'Astrophysique de Paris, UMR7095, CNRS, Universit\'e Paris VI, 98bis Boulevard Arago, Paris, France \label{IAP}, \and 
   Laboratoire de Chimie et de Physique Quantiques, Universit\'e de Toulouse (UPS) and CNRS, 118 route de Narbonne, F-31400 Toulouse, France \label{Toulouse}\and 
   Department of Astronomy and Carl Sagan Institute, Cornell University, 122 Sciences Drive, Ithaca, NY, 14853, USA\label{Cornell}\and
   Met Office, Fitzroy Road, Exeter, EX1 3PB, UK\label{Met}}

   \date{Received December 20, 2019 / Accepted March 11, 2020}

 
  \abstract{We present a new set of solar metallicity atmosphere and evolutionary models for very cool brown dwarfs and self-luminous giant exoplanets, which we term \texttt{ATMO} 2020. Atmosphere models are generated with our state-of-the-art 1D radiative-convective equilibrium code \texttt{ATMO}, and are used as surface boundary conditions to calculate the interior structure and evolution of $0.001-0.075\,\mathrm{M_{\odot}}$ objects. Our models include several key improvements to the input physics used in previous models available in the literature. Most notably, the use of a new H--He equation of state including ab initio quantum molecular dynamics calculations has raised the mass by $\sim1-2\%$ at the stellar--substellar boundary and has altered the cooling tracks around the hydrogen and deuterium burning minimum masses. A second key improvement concerns updated molecular opacities in our atmosphere model \texttt{ATMO}, which now contains significantly more line transitions required to accurately capture the opacity in these hot atmospheres. This leads to warmer atmospheric temperature structures, further changing the cooling curves and predicted emission spectra of substellar objects. We present significant improvement for the treatment of the collisionally broadened potassium resonance doublet, and highlight the importance of these lines in shaping the red-optical and near-infrared spectrum of brown dwarfs. We generate three different grids of model simulations, one using equilibrium chemistry and two using non-equilibrium chemistry due to vertical mixing, all three computed self-consistently with the pressure-temperature structure of the atmosphere. We show the impact of vertical mixing on emission spectra and in  colour-magnitude diagrams, highlighting how the $3.5-5.5\,\mathrm{\mu m}$ flux window can be used to calibrate vertical mixing in cool T--Y spectral type objects.}


    \titlerunning{New Models for Brown Dwarfs}
\authorrunning{Phillips et al.}

   \maketitle
%

\section{Introduction} \label{sec:intro}

 \def\simle{\,\hbox{\hbox{$ < $}\kern -0.6em \lower 1.0ex\hbox{$\sim$}}\,}

\def\simgr{\,\hbox{\hbox{$ > $}\kern -0.6em \lower 1.0ex\hbox{$\sim$}}\,}

The absence or lack of steady hydrogen fusion in the cores of brown dwarfs means that these objects cool over time by radiating away their internal thermal energy. This cooling leads to a degeneracy in mass, age, effective temperature, and luminosity, making the fundamental properties of brown dwarfs, particularly isolated field objects, difficult to determine. The rate at which these objects cool is regulated by the atmosphere, which imprints its complex and changing chemical composition of molecules and condensate species onto the emitted radiation, forming the M-L-T-Y spectral sequence \citep{Kirkpatrick_2005, Helling_Casewell_2014}. A reliable model of the atmosphere and its evolution over time therefore lies at the core of our understanding of brown dwarfs and substellar objects. Illustrative of this, the fundamental properties of brown dwarfs are often obtained by fitting synthetic spectra from grids of atmosphere models and then inferring the mass and age of the object using evolution models (e.g. \citet{Saumon_2006, Saumon_2007, Burningham_2011, Leggett_2019}). Understanding the atmospheres of brown dwarfs has further motivation since the physics, chemistry, and composition is shared with hot Jupiters and directly imaged exoplanets \citep{Burrows_2001}, meaning that useful analogies can be drawn between these objects.

Traditionally, the atmospheres of brown dwarfs and giant planets are modelled with 1D codes which solve for the atmospheric temperature structure in radiative-convective flux balance \citep{Marley_2015, Fortney_2018}. These codes are used to compute grids of models spanning effective temperature and surface gravity containing temperature structures and top of the atmosphere emission spectra for comparison to observations. These atmosphere structures are then coupled as non-grey  surface boundary conditions \citep{Chabrier_Baraffe_1997} to interior structure models to compute the cooling and evolution over time. Two of the earliest model sets that follow this framework and that are widely used in the literature include \citet{Burrows_1997} and the AMES-Cond models of \citet{Baraffe_2003} (hereafter B03). \citet{Saumon_2008} (hereafter SM08) presented coupled atmosphere and evolutionary calculations, additionally varying the cloud sedimentation efficiency \citep{Ackerman_Marley_2001} within their atmospheric outer boundary condition, in order to investigate the impact of clouds on brown dwarf evolution. More recently, \citet{Fernandes_2019} used existing atmosphere models in the literature as surface boundary conditions to a stellar evolution code to  investigate the effects of including additional metals in the interior equation of state (EOS) on the substellar boundary.

Beyond these coupled atmosphere and evolution models  numerous improvements and complexities have been added to 1D atmosphere codes in an  attempt to reproduce and explain various features of the observed brown dwarf cooling sequence. Cloud models have been developed \citep{Allard_2001, Ackerman_Marley_2001, Helling_2008} and invoked to explain the reddening L dwarf spectral sequence \citep{Chabrier_2000, Cushing_2008, Stephens_2009, Witte_2011}, the sharp change to bluer near-infrared colours at the L-T transition \citep{Allard_2001, Burrows_2006, Marley_2010, Charnay_2018}, and the reddening observed in the spectra of late T and Y dwarfs \citep{Morley_2012, Morley_2014}. A reduction in the atmospheric temperature gradient has also been explored in 1D models to provide an alternative explanation to the cloudy scenario for this reddening observed along the cooling sequence \citep{Tremblin_2015, Tremblin_2016, Tremblin_2017b}. This reduction in the temperature gradient has been linked to diabatic convection triggered by the $\mathrm{CO/CH_4}$ transition in brown dwarf atmospheres \citep{Tremblin_2019}. 

Along with these additional complexities, there has also been significant improvement in the fundamental input physics to 1D atmosphere models. The opacity for important molecular absorbers has improved through more complete high-temperature line lists \citep{Tennyson_2018}, which has altered the temperature structures and synthetic emission spectra in 1D model grids \citep{Saumon_2012, Malik_2019}. There has been significant theoretical improvement in the pressure broadened line shapes of the alkali metals Na and K \citep{Allard_2016, Allard_2019}, which shape the red-optical and near-infrared spectra of brown dwarfs. Non-equilibrium chemistry due to vertical mixing is a prevalent feature in brown dwarf observations \citep{Noll_1997, Saumon_2000, Saumon_2006, Geballe_2009, Leggett_2015, Leggett_2017}, and theoretical studies have improved our understanding of the impact of such processes in 1D models \citep{Hubeny_2007, Zahnle_Marley_2014, Tremblin_2015, Drummond_2016}. To further aid in the study of non-equilibrium chemistry processes, complex chemical kinetics networks containing thousands of reactions between important molecules in exoplanet and brown dwarf atmospheres have been developed and refined \citep{Moses_2011, Venot_2012, Tsai_2017, Tsai_2018, Venot_2019}.


Along with these theoretical improvements, the  study  of  brown  dwarfs  and  giant  exoplanets is being driven by ever-improving instrumentation that is becoming sensitive to cooler objects. Over the last decade the \textit{WISE} mission \citep{Wright_2010} has uncovered the coolest spectral type (known as the Y dwarfs) \citep{Cushing_2011, Kirkpatrick_2012}, including the coldest known brown dwarf at $T_{\mathrm{eff}}\sim250\,\mathrm{K}$ just $2\,$pc from the sun \citep{Luhman_2014}. At a few times warmer than Jupiter, these objects provide excellent analogues for Jovian-like worlds outside of our solar system, and are proving challenging for atmosphere models \citep{Morley_2018, Leggett_2019}. Ongoing projects are likely to discover more objects in this temperature range (e.g. \citet{Marocco_2019}). Accurate and reliable atmosphere and evolutionary models are important for placing mass and age constraints on these newly discovered objects, understanding the rich chemistry and physics taking place in their atmospheres, and determining the low-mass end of the initial mass function \citep{Kirkpatrick_2019}.

In this work we present a new set of coupled atmosphere and evolutionary models for brown dwarfs and giant exoplanets. This grid, which we name $\texttt{ATMO 2020}$, includes numerous improvements to the input physics for modelling substellar objects, and thus supersedes the widely used AMES-Cond grid of B03. We use our 1D atmosphere code $\texttt{ATMO}$ to generate self-consistent models with equilibrium chemistry and non-equilibrium chemistry due to vertical mixing. We include updated line lists for important molecular absorbers and improved line shapes for the collisionally broadened potassium resonance lines. Finally, we couple these atmosphere models to an interior structure model which uses a new H--He  EOS from \citet{Chabrier_2019} including ab initio quantum molecular dynamics calculations. 


The paper is organised as follows. In Section \ref{sec:methods} we outline the details of the grid and the tools used to generate the models. In Section \ref{sec:K_broadening} we present the impact of including new potassium resonant line shapes from \citet{Allard_2016} (hereafter A16) in our 1D atmosphere model $\texttt{ATMO}$, and compare them to other line shapes available in the literature. Our main results are presented in Section \ref{sec:Results}, where we show how modelling improvements have impacted the predicted cooling tracks, emission spectra, and colours of substellar objects by comparison to other model grids and observational datasets. Finally, we discuss and summarise our work in Section \ref{sec:discussion}.

\section{Grid set-up and methods} \label{sec:methods}
 
\subsection{ Model grid} \label{sec:grid}

The model set consists of a grid of solar metallicity atmosphere models spanning $T_{\mathrm{eff}}=200-3000\,\mathrm{K}$ and $\log (g)=2.5-5.5$ ($g$ in cgs units), with steps of 100\,K for $T_{\mathrm{eff}}>600\,\mathrm{K}$,  50\,K for $T_{\mathrm{eff}}<600\,\mathrm{K}$, and   0.5 in $\log(g)$.  We note that we extend our grid of models to $T_{\mathrm{eff}}=3000\,\mathrm{K}$ in order to follow the evolution of the most massive brown dwarfs from very early stages starting from hot luminous initial models. However, the range of validity of our atmosphere models is $T_{\mathrm{eff}}\simle2000\,\mathrm{K}$ since we do not include some sources of opacity (e.g. some hydrides and condensates) that form at higher temperatures (see Sections \ref{sec:opacity} and \ref{sec:discussion}).

We generate three atmosphere grids with different chemistry schemes spanning this parameter range. The first is calculated assuming chemical equilibrium, and the second and third are calculated assuming non-equilibrium chemistry due to vertical mixing with different mixing strengths. Each model in each of the grids is generated with the \texttt{ATMO} code (see Section \ref{sec:ATMO}), and consists of a pressure-temperature (P-T) profile, chemical abundance profiles, and a spectrum of the emergent flux at the top of the atmosphere. These models are publicly available for download\footnote{\url{http://opendata.erc-atmo.eu}\label{note1}}\footnote{ \url{http://perso.ens-lyon.fr/isabelle.baraffe/ATMO2020/}.\label{note2}}.

The P-T profiles from the model atmosphere grid are then used as outer boundary conditions for the interior structure model to follow the evolution of $0.001-0.075\,\mathrm{M_{\odot}}$ objects from $0.001-10\,\mathrm{Gyr}$. We follow the evolution of the object's effective temperature, luminosity, radius, gravity, and absolute magnitudes in a range of photometric filters. Absolute magnitudes are derived by calculating the flux density in a given photometric filter for each spectrum in the atmosphere grid. The flux density can then be interpolated to the $T_{\mathrm{eff}}$ and $\log(g)$ for a given mass and age, and the corresponding radius used to compute the absolute magnitude. The zero point is calculated from a Vega spectrum.  The evolutionary tracks for a given mass are also publicly available for download\textsuperscript{\ref{note1}}\textsuperscript{\ref{note2}}. We  provide more detail on our atmosphere code, chemistry schemes, opacity database, and interior structure model in Sections \ref{sec:ATMO}, \ref{sec:chem}, \ref{sec:opacity}, and \ref{sec:interior_structure}, respectively. 

\subsection{One-dimensional atmosphere model - \texttt{ATMO}} \label{sec:ATMO}

\texttt{ATMO} is a 1D--2D atmosphere model developed to study hot Jupiters \citep{Amundsen_2014, Drummond_2016, Tremblin_2017a, Goyal_2018, Goyal_2019, Drummond_2019} and brown dwarfs \citep{Tremblin_2015, Tremblin_2016, Tremblin_2017b}. The model (in 1D) solves for the P-T structure of an atmosphere that is self-consistent with radiative-convective flux balance for a given internal heat flux, and hydrostatic equilibrium for a given surface gravity. This type of model, often termed a radiative-convective equilibrium model, has a long history of being used to study brown dwarf and giant planet atmospheres, and we refer the reader to \citet{Marley_2015} for a thorough review of these models in this context.

The P-T structure is solved by \texttt{ATMO} on a logarithmic optical depth grid defined in the spectral band between 1.20 and $1.33\,\mu m$. We use 100 model levels, with the outer boundary condition in the first model level fixed at a pressure of $10^{-5}\,$bar and given an optical depth of $\tau\sim10^{-4}-10^{-7}$ depending on $\log (g)$. The inner boundary condition in the last model level is not fixed in pressure and given an optical depth of $\tau=1000$. A first guess of pressure and temperature is assigned to each model level, and then the model iterates the P-T structure towards radiative-convective and hydrostatic equilibrium using a Newton-Raphson solver. On each iteration chemical abundances are calculated for the current P-T structure, opacities are obtained from pre-computed look-up tables for individual gases, and the radiative and convective fluxes are calculated. The P-T structure is generally considered converged when radiative-convective flux balance and hydrostatic equilibrium is satisfied to an accuracy of $\le1\times10^{-3}$  in each model level.

\texttt{ATMO} can calculate chemical abundances assuming thermodynamic equilibrium or assuming non-equilibrium chemistry due to vertical mixing in the atmosphere. The chemistry schemes used in this work are discussed in Section \ref{sec:chem}. Once the chemical abundances have been computed, the opacities used by \texttt{ATMO} are loaded from pre-computed correlated-$k$ tables for individual gases (discussed in Section \ref{sec:opacity}), and are combined within the code using the random overlap method with resorting and rebinning to get the total mixture opacity \citep{Amundsen_2017}. This method ensures the opacities are completely consistent with the pressure, temperature, and abundances on every iteration. 

The radiative flux is computed by solving the integral form of the radiative transfer equation in 1D plane-parallel geometry following \citet{Bueno_1995}. We include isotropic scattering and sample 16 ray directions with a discrete ordinate method using Gauss-Legendre quadrature. The convective flux is computed using mixing length theory \citep{Henyey_1965} using the same method as \citet{Gustafsson_2008}, with a mixing length of 2 times the local pressure scale height. The adiabatic gradient is computed using EOS tables from \citet{Saumon_1995}.

\subsection{Chemistry schemes} \label{sec:chem}

Chemical equilibrium abundances are calculated using a Gibbs energy minimisation scheme based on that of \citet{Gordon1994}. We use 76 gas phase species (including ionic species) and 92 condensate species with thermodynamic data from \citet{McBride_1993, McBride_2002}\footnote{The full list of species is available on \url{http://opendata.erc-atmo.eu} and \url{http://perso.ens-lyon.fr/isabelle.baraffe/ATMO2020/}.}. To form these species we include 23 elements: H, He, C, N, O, Na, K, Si, Ar, Ti, V, S, Cl, Mg, Al, Ca, Fe, Cr, Li, Cs, Rb, F, and P. The present models adopt the solar composition of \citet{Asplund_2009} with revisions of the elemental abundances of C, N, O, P, S, K, and Fe from the CIFIST project \citep{Caffau_2011}. Our equilibrium chemistry scheme has been benchmarked against the \texttt{GGchem} code \citep{Woitke_2018} in \citet{Goyal_2019_erratum}, and against the \texttt{Exo-REM} and \texttt{petitCODE} 1D atmosphere models in \citet{Baudino_2017}.

We adopt the rainout approach for the treatment of condensates as described in \citet{Goyal_2019_erratum} whereby once a condensate forms, the elements comprising that condensate will be depleted from the current level and all the model levels above (lower pressures). This approach models the settling or sinking of cloud particles in an atmosphere which depletes elemental abundances at lower pressures \citep{Burrows_Sharp_1999}. Evidence for this rainout process has been found in the retrieved abundances of alkali metals for late T and Y dwarfs \citep{Line_2017, Zaleksy_2019}. 

To calculate non-equilibrium chemical abundances we have implemented the chemical relaxation scheme of \citet{Tsai_2018}. Chemical relaxation schemes take the approach of relaxing a species back to its equilibrium abundance, following a perturbation, on a given timescale. The chemical timescale for each species is estimated or parametrised based on a complex chemical kinetics network \citep{Cooper_Showman_2006, Zahnle_Marley_2014, Tsai_2018}. \citet{Tsai_2018} find the rate-limiting reactions within a chemical network to derive relaxation timescales of $\mathrm{H_2O}$, CO, $\mathrm{CO_2}$, $\mathrm{CH_4}$, $\mathrm{N_2}$, and $\mathrm{NH_3}$ over several P-T regimes in the range 500 to 3000\,K, and 0.1\,mbar to 1\,kbar. 

We choose to adopt this chemical relaxation scheme over full chemical kinetics networks for computational efficiency and consistent convergence throughout the grid when solving for a self-consistent P-T profile. The relaxation method is more computationally efficient as it avoids the need to solve the large, stiff system of ordinary differential equations needed when using full chemical kinetics networks. The P-T profile is reconverged on the fly while integrating over time for the non-equilibrium abundances every 50 iterations of the numerical solver. Reconverging the profile more often than every 50 iterations gives negligible differences in the final P-T structure, abundances, and emission spectrum. The chemistry is integrated for a minimum of $1\times10^{10}\,\mathrm{s}$, and is considered converged and in a steady state when $dn/n<1\times10^{-2}$ and $(dn/n)/dt<1\times10^{-4}$ for all species, where $n$ is the species number density. This self-consistent non-equilibrium chemistry approach is similar to that used in hot Jupiter models presented in \citet{Drummond_2016}.

\begin{figure}[t!]
\centering
\resizebox{1.0\hsize}{!}{\includegraphics{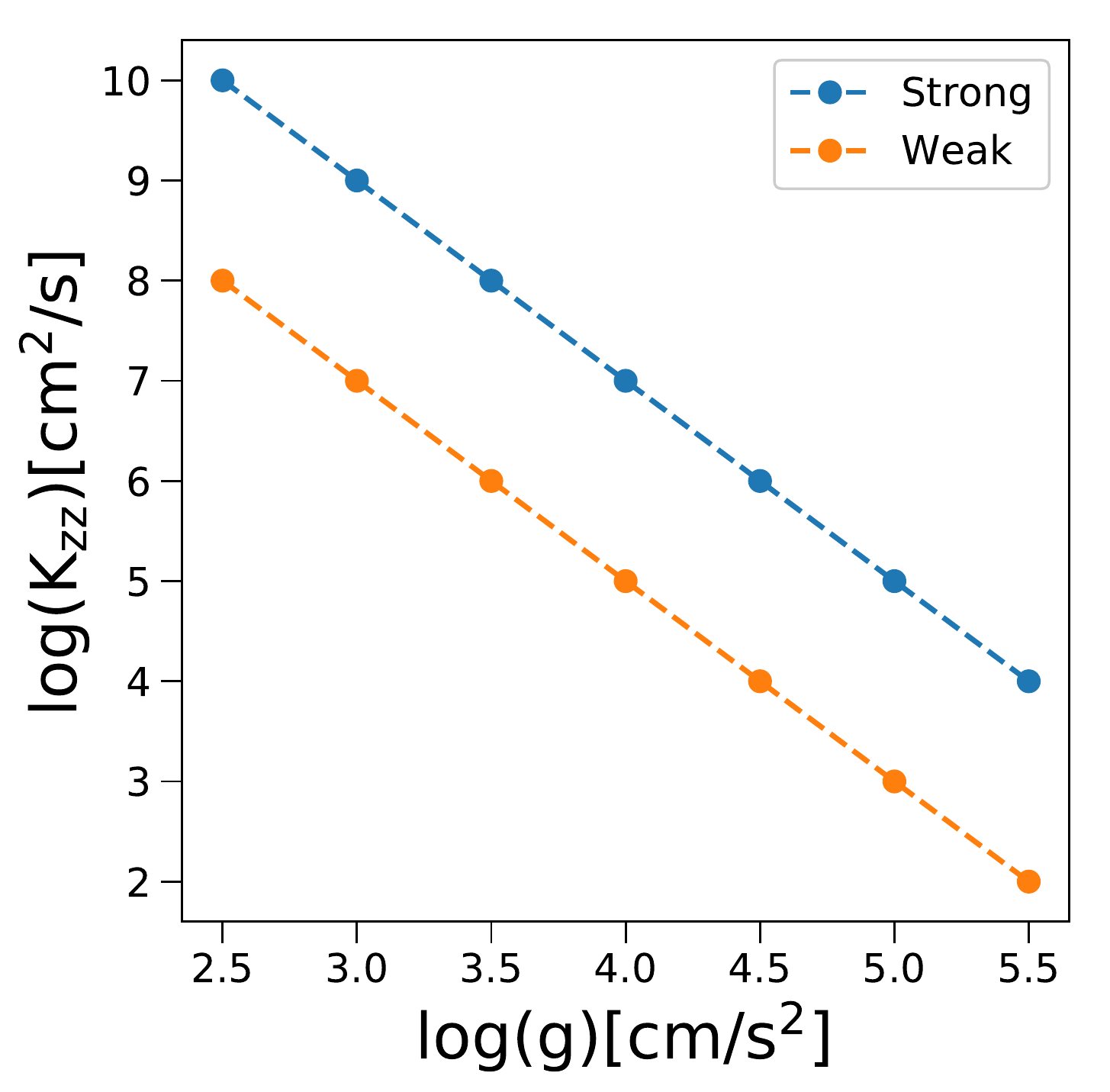}}
\caption{Vertical mixing relationships  with surface gravity (strong and weak; see text)  used in the generation of non-equilibrium atmosphere models in this work. \label{fig:Kzz_plot}}
\end{figure}

Vertical mixing in the atmosphere is parametrised using the eddy diffusion coefficient $K_{zz}$ in $\mathrm{cm^2s^{-1}}$, and is assumed to be constant throughout the atmosphere. We scale the eddy diffusion coefficient with surface gravity since the typical dynamical timescale $t$ can be approximated as

\begin{equation}
    t\sim\frac{H_P^2}{K_{zz}}\propto\frac{1}{g^2K_{zz}},
\end{equation}

\noindent where $H_P$ is the atmospheric scale height. Within this approximation, we keep the dynamical timescale $t$ constant by changing the value of $K_{zz}$ by an order of magnitude for a $\log(g)$ step of 0.5 within the grid. We generate atmosphere model grids with two $K_{zz}$ scaling relationships with surface gravity as shown in Figure \ref{fig:Kzz_plot};  we refer to these relationships as `strong' and `weak' mixing throughout this work. 

Our choice of mixing strengths come from approximate values in the literature which have been found to provide reasonable comparisons to observations of late T and Y dwarfs. For example, \citet{Leggett_2017} found $K_{zz}$ values in the range $10^4$-$10^6\,\mathrm{cm^2/s}$ provided reasonable comparison to the $[4.5]$-$M$ colours of late T and Y dwarfs for model sequences with a constant gravity of $\log(g)=4.5$ (see their Figure 7). We have therefore adopted to set $\log(K_{zz})=4$ and $\log(K_{zz})=6$ in the `weak' and `strong' cases respectively at $\log(g)=4.5$ and scale $K_{zz}$ with gravity.

We note that $K_{zz}$ has often been estimated by assuming it is the same diffusion coefficient as that derived from mixing length theory of convection, i.e. $D_{\mathrm{mix}} \sim l_{\mathrm{mix}}v_{\mathrm{mlt}}$, with $l_{\mathrm{mix}}$ the mixing length and $v_{\mathrm{mlt}}$ the convective velocity \citep{Gierasch_Conrath_1985, Ackerman_Marley_2001}. This however has to be extrapolated to the convectively stable radiative regions of the atmosphere where a number of complex processes such as gravity waves and convective overshooting \citep{Freytag_1996, Kupka_2018} may drive the mixing. The value of $K_{zz}$ has also been approximated from 3D numerical simulations of hot Jupiters including passive tracer transport \citep{Parmentier_2013, Zhang_Showman_2018b}. These approaches to estimating $K_{zz}$ have their limitations and none has provided a quantitative picture that has reached a consensus in the community. In this work we therefore choose to adopt a simpler approximation for $K_{zz}$ to examine the trends of non-equilibrium chemistry in colour-magnitude diagrams (Section \ref{sec:CMDs}), and we leave more sophisticated studies of $K_{zz}$ for future work.


\subsection{Opacity database}\label{sec:opacity}

Our opacity database used by \texttt{ATMO} consists of 22 atomic and molecular species shown in Table \ref{tab:opacity}, and our methodology for calculating these opacities is presented in detail in \citet{Amundsen_2014}. The absorption coefficient is calculated on a wavenumber grid spanning $0-50000\,\mathrm{cm^{-1}}$ and a resolution of $0.001\,\mathrm{cm^{-1}}$, with transitions from the line list sources provided in Table \ref{tab:opacity}. Each line is broadened including both Doppler and pressure broadening with collisions from $\mathrm{H_2}$ and He (wherever data is available for each perturber), the dominant species in brown dwarf and hot Jupiter atmospheres. This is done on a pressure and temperature grid, with 40 logarithmically spaced pressure points from $10^{-9}-10^{3}\,$bar, and 20 logarithmically spaced temperature points in the range $70-3000\,$K. Pressure broadening parameters for $\mathrm{H_2}$ and He are often not provided in the line lists given in Table \ref{tab:opacity}, and are therefore obtained from alternative sources found in Table C1 of \citet{Goyal_2018}. The implementation of these pressure broadening parameters and  our numerical considerations regarding line wing cutoffs are discussed in \citet{Amundsen_2014}.

\begin{table}[b]
\caption{Opacity database used by \texttt{ATMO}.} 
\label{tab:opacity}  
\centering     
\begin{tabular}{c c}
\hline\hline                        
Species & Source \\  
\hline                                   
$\mathrm{H_2}$-$\mathrm{H_2}$, $\mathrm{H_2}$-$\mathrm{He}$ & \citet{Richard_2012} \\
$\mathrm{H^-}$ & \citet{John_1988} \\
$\mathrm{H_2O}$ & \citet{Barber_2006} \\
$\mathrm{CO_2}$ & \citet{Tashkun_2011} \\
$\mathrm{CO}$ & \citet{Rothman_2010} \\
$\mathrm{CH_4}$ & \citet{Yurchenko_2014} \\
$\mathrm{NH_3}$ & \citet{Yurchenko_2011} \\
$\mathrm{Na}, \mathrm{K}, \mathrm{Li}, \mathrm{Rb}, \mathrm{Cs}, \mathrm{Fe}$ & VALD \citep{Heiter_2015} \\
$\mathrm{TiO}$ & \citet{Plez_1998} \\
$\mathrm{VO}$ & \citet{McKemmish_2016} \\
$\mathrm{FeH}$ & \citet{Wende_2010} \\
$\mathrm{PH_3}$ & \citet{Sousa-Silva_2015} \\
$\mathrm{HCN}$ & \citet{Barber_2014} \\
$\mathrm{C_2H_2}$, $\mathrm{H_2S}$ & \citet{Rothman_2013} \\
$\mathrm{SO_2}$ & \citet{Underwood_2016} \\
\hline                                             
\end{tabular}
\end{table}

Unlike the molecular line lists used in our database, the VALD line lists for the atomic species contain van der Waals coefficients, which can be used to calculate pressure broadened line widths \citep{Sharp_Burrows_2007}. These coefficients are used to calculate widths for all lines except for the Na and K resonance doublets located at  $\sim0.59\,\mu m$ and $\sim0.77\,\mu m$, respectively, for which a Lorentzian line profile has been shown to be insufficient \citep{Burrows_2000, Burrows_2002}. The high pressures and temperatures in brown dwarf atmospheres cause these resonance lines to be broadened up to $\sim4000\,\mathrm{cm^{-1}}$ away from the line centre, and more detailed calculations of these line shapes beyond a Lorentzian profile are required. In previous works with \texttt{ATMO}, we  used Na and K line shapes from both \citet{Burrows_Volobuyev_2003} and \citet{AllardN_2007} (hereafter BV03 and A07 respectively). For this work we have updated our K resonant line shapes with those presented in A16. We discuss this in more detail in Section \ref{sec:K_broadening}, and  compare the effect that different line shape calculations can have on model brown dwarf atmospheres and synthetic observations. 

The radiative transfer equation in \texttt{ATMO} can be solved at the native resolution (0.001$\,\mathrm{cm}^{-1}$) of the absorption cross sections (commonly known as the line-by-line approach); however, this is computationally expensive, particularly when iterating for a consistent P-T structure. The more computationally efficient correlated-$k$ approximation is therefore used \citep{Lacis_Oinas_1991}, which is an approach widely adopted by both the Earth atmosphere and exoplanet communities. The open source UK Met Office radiative transfer code SOCRATES \citep{Edwards_1996, Edwards_Slingo_1996} is used to generate correlated-$k$ opacity tables for each species in our database, and our methodology is described and tested in \citet{Amundsen_2014}. These tables are computed on the same P-T grid as the full resolution absorption coefficient files, and are provided at 32-, 500-, and 5000-band spectral resolutions. The spacing in the 32-band files is as shown in Table 4 of \citet{Amundsen_2014}, and these tables are used when iterating for a consistent P-T structure with \texttt{ATMO}, improving computational efficiency while maintaining an accurate heating rate. The 500 and 5000 bands are evenly spaced in wavenumber between 1 and 50000$\,\mathrm{cm^{-1}}$, and the 5000-band tables are used to generate emission spectra shown in this work.

\subsection{Interior structure and evolution model} \label{sec:interior_structure}

Calculations of interior structure and evolutionary models are based on the Lyon stellar evolution code, and are described in detail in our previous works \citep{Chabrier_Baraffe_1997, Baraffe_1998, Baraffe_2003}. The structure models are based on the coupling between interior profiles and the chemical equilibrium atmospheric structures described previously at an optical depth $\tau$ = 1000. We note that this is deeper than our previous models which used $\tau=100$ to couple the atmosphere to the interior. However, the radial extension of the atmosphere at $\tau=1000$ is still negligible compared to the total radius of the object, and thus the Stefan-Boltzmann condition ($L=4\pi\sigma R^2 T_{\mathrm{eff}}^4$) is still satisfied. We use a solar metallicity helium mass fraction $Y=0.275$ \citep{Asplund_2009} to be consistent with our previous models \citep{Baraffe_1998, Chabrier_2000, Baraffe_2003}. Since we are using a metal-free EOS, the presence of metals with mass fraction $Z$ can be mimicked by an equivalent He mass fraction $Y_{\mathrm{eq}}=Y+Z$ \citep{Chabrier_Baraffe_1997}. We use $Z=0.0169$ giving $Y_{\mathrm{eq}}=0.2919$.

The main change in terms of inner structure input physics concerns the EOS. 
In this work we use the new EOS for H--He mixtures presented by \citet{Chabrier_2019}, which includes ab initio quantum molecular dynamics calculations in the regime of pressure dissociation and ionisation. This is a significant improvement over the semi-analytic H--He  EOS of \citet{Saumon_1995} (SCVH) used in this regime in all our previous models \citep{Baraffe_1998, Chabrier_2000, Baraffe_2003}. 

For the sake of comparison, we have also computed a set of evolutionary models with the SCVH EOS to determine the impact of the new EOS. We note that the SCVH EOS is used in the atmosphere models (see Section \ref{sec:ATMO}). There is, however, no difference between the SCVH EOS and the new EOS of \citet{Chabrier_2019} in the atmospheric P-T regime, which is close to a perfect gas. There is thus no inconsistency when using the SCVH EOS in the atmosphere models and the new EOS in the inner structure models.

\section{Potassium broadening}\label{sec:K_broadening}

The alkali metals sodium Na and potassium K play a key role in brown dwarf atmospheres. They are abundant in the gas phase until they condense into KCl and $\mathrm{Na_2S}$ \citep{Lodders_1999}, and have strong resonance lines at $\sim0.59\,\mu m$ and $\sim0.77\,\mu m$, respectively, that are present in late L and T dwarfs \citep{Kirkpatrick_1999, Burgasser_2003}. The line shapes are determined by the potential field of $\mathrm{H_2}$ perturbing the ground and excited states of the alkali atom, and in brown dwarf atmospheres these resonance lines become broadened out to thousands of angstroms away from the line core, shaping the visible and red-optical spectra of cool brown dwarfs. As such, Lorentzian line profiles are not sufficient to model the collisional broadening effects on these alkali metals \citep{Allard_1982, Burrows_2000, Burrows_2002, Allard_2019}, and more detailed quantum chemical calculations of the interaction potentials of these collisions are required to accurately model Na and K line shapes. 

\begin{figure}[b!]
\centering
\resizebox{1.0\hsize}{!}{\includegraphics{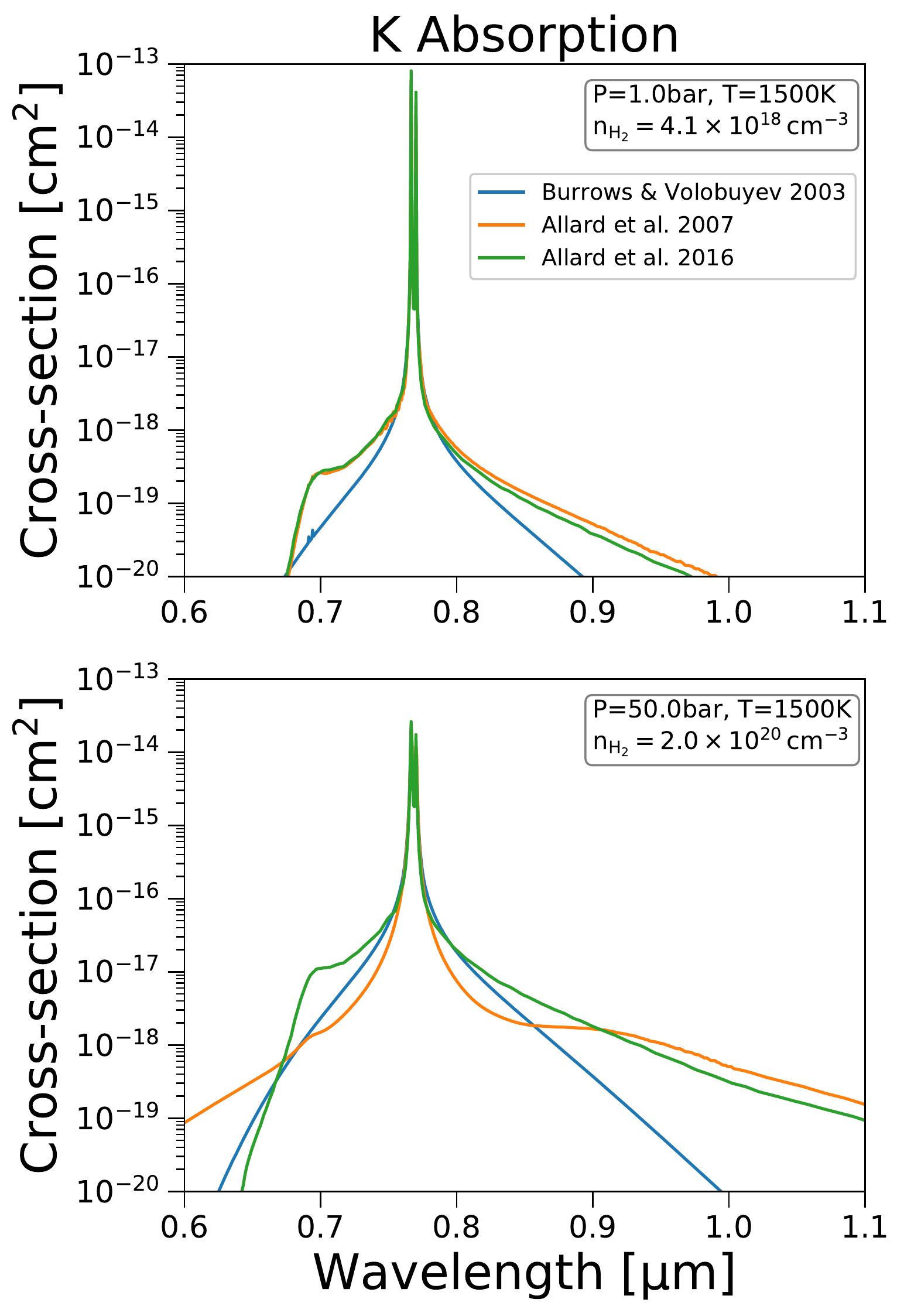}}
\caption{Absorption cross section of potassium calculated with different broadening treatments for the D1 and D2 resonance doublet, at a pressure of 1 bar and a temperature of 1500K (top panel) and a pressure of 50 bar and a temperature of 1500K (bottom panel).\label{fig:K_opacity}}
\end{figure}

\begin{figure*}[t!]
\centering
\resizebox{\hsize}{!}{\includegraphics{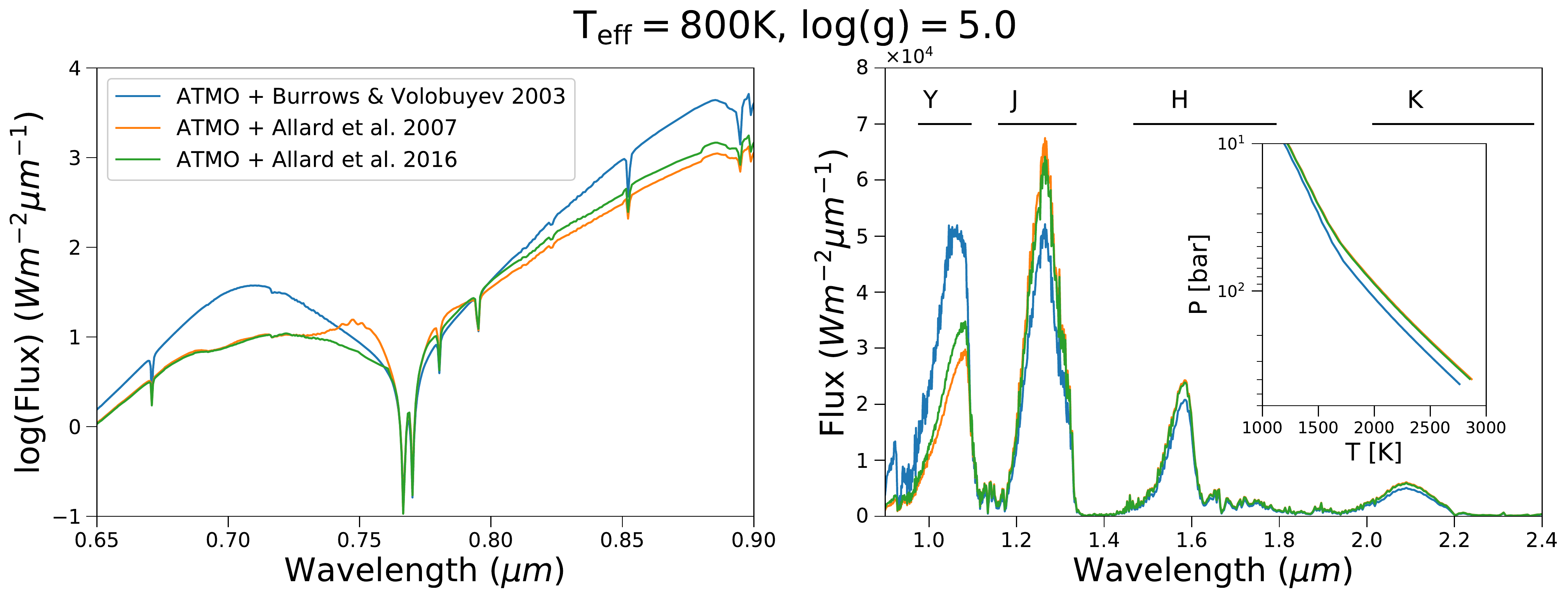}}
\caption{Comparison of the emission spectra from the top of the atmosphere for $T_{\mathrm{eff}}=800\mathrm{K}$ and $\mathrm{\log(g)=5.0}$ calculated with alkali broadening from BV03; A07; A16. Indicated in the right  plot are the locations of the Mauna Kea near-infrared filters. \label{fig:K_spec}}
\end{figure*}

Both BV03 and A07 have presented alkali broadening calculations which can be used in 1D radiative-convective models of brown dwarfs and exoplanets. BV03 calculate the interaction potentials of the ground and excited states of Na and K perturbed by $\mathrm{H_2}$ and He as a function of distance and orientation angle. Using these potentials BV03 computed absorption line profiles using the Franck-Condon model in the quasi-static limit out to thousands of angstroms from the line core. A07 used valence pseudopotentials to compute molecular potentials of Na and K perturbed by $\mathrm{H_2}$ and He, and used the semi-classical unified line shape theory of \citet{Allard_1999} to calculate the collisional profiles of the Na- and K-$\mathrm{H_2}$ resonance lines.

Previous works with \texttt{ATMO} have used both the BV03 and A07 broadening treatments as it remains unclear which performs best when reproducing observations. The BV03 profiles used in \texttt{ATMO} are implemented by \citet{Baudino_2015}. \citet{Baudino_2017} benchmarked the BV03 and A07 alkali broadening schemes in a 1D radiative-convective model showing large uncertainties in the predicted transmission spectra of hot Jupiters and the emission spectra of brown dwarfs. When generating the grid of brown dwarf atmosphere models in this work we found similar uncertainties. In particular, the differences in opacity in the far red wing of the K doublet cause substantial differences in the predicted near-infrared spectra where the peak in brown dwarf emission lies (see Figure \ref{fig:K_spec}). This motivated us to implement the new K resonance line profiles presented in A16.

The A16 line profiles follow the same framework as A07, with improvements on the determination of the intermediate- and long-range part of the $\mathrm{K-H_2}$ potential and the inclusion of spin-orbit coupling. The wing profiles of A16 are tabulated for temperatures between 600 and $3000\,$K as powers of density expansion \citep{Allard_1999}. The new line profiles in A16 are valid for $n_{\mathrm{H_2}}<10^{21}\,$cm$^{-3}$, whereas the A07 profiles were valid for  $n_{\mathrm{H_2}}<10^{19}\,\mathrm{cm^{-3}}$, where $n_{\mathrm{H_2}}$ is the number density of $\mathrm{H_2}$.

In Figure \ref{fig:K_opacity} we show the absorption cross section  of potassium employed in \texttt{ATMO} using broadening schemes from BV03, A07, and A16, at pressures and temperatures typical of the red-optical to near-infrared photosphere of T-type brown dwarfs. The top panel displays the K opacity for P=1\,bar and T=1500\,K. This corresponds to a $n_{\mathrm{H_2}}<10^{19}\,\mathrm{cm^{-3}}$ regime within which both the A07 and A16 profiles are valid. Therefore, the A07 and A16 wing profiles predict a similar strength quasi-molecular $\mathrm{K-H_2}$ line satellite in the blue wing at $\sim0.7\,\mu m$, which is not captured by the BV03 wing profiles. The bottom panel of Figure \ref{fig:K_opacity} shows the K absorption cross section at a higher pressure of 50\,bar corresponding to a $10^{19}\,\mathrm{cm^{-3}}<n_{\mathrm{H_2}}<10^{21}\,\mathrm{cm^{-3}}$ regime within which the A07 tables are no longer valid, while the A16 profiles are. The A07 profiles therefore predict a much weaker line satellite than the A16 profiles. At both 1 and 50\,bar, the opacity differs considerably in the red wing at $\sim1\,\mu m$, with the BV03 profiles giving significantly less absorption than the A07 and A16 profiles.

In Figure \ref{fig:K_spec} we show a synthetic red-optical and near-infrared emission spectrum of a $T_{\mathrm{eff}}=800\,\mathrm{K}$, $\log(g)=5.0$ T-type brown dwarf calculated with the BV03, A07, and A16 broadening schemes. The red-optical spectra in the left panel shows the difference in the emission around the potassium D1 and D2 resonance doublet. There is a noticeable difference between the emission in the blue wing around $\sim0.7\,\mu m$ due to the $\mathrm{K-H_2}$ quasi-molecular feature predicted by A07 and A16 compared to BV03. The lower absorption in the red wing in the BV03 case leads to more flux emerging through the $Y$ band at $\sim1\,\mu m$ compared to the A07 and A16 cases. The large differences in opacity in the BV03 profiles compared to the A07 and A16 profiles also causes differences in the temperature profile when reconverging the atmospheric structure to find radiative-convective equilibrium. P-T profiles generated including BV03 alkali opacity are several hundred Kelvin cooler for pressures above 5\,bar than profiles generated with A07 and A16 opacity. This leads to the redistribution of flux across the near-infrared seen in the  right  panel of Figure \ref{fig:K_spec}. We note that this flux distribution only occurs if the model is generated self-consistently with a reconverged P-T structure when switching between opacity sources.

\section{Results}\label{sec:Results}

This section presents our main results and is organised as follows. In Section \ref{sec:evol_tracks} we present our new substellar evolutionary tracks; we highlight the impact of the new EOS, and compare it to other calculations in the literature. In Section \ref{sec:dyn_masses} we compare these new evolutionary tracks to dynamical mass measurements of brown dwarfs. We demonstrate the impact of non-equilibrium chemistry due to vertical mixing on synthetic emission spectra in Section \ref{sec:NEQ_spec}. In Section \ref{sec:CMDs} we compare our new models to other models and observational datasets in colour-magnitude diagrams. In Section \ref{sec:spectra} we make spectral comparisons to other models to highlight improvements in the atmospheric opacities. Finally, in Section \ref{sec:obs_spectra} we make initial comparisons of our new models to the observed spectra of cool brown dwarfs across the T-Y transition.
\subsection{Evolutionary tracks}\label{sec:evol_tracks}

In this section we present and compare the new set of atmosphere models and evolutionary tracks to others in the literature in order to highlight model improvements. We choose two families of brown dwarf models that are widely used in the community for comparison, the Lyon group and the Saumon \& Marley group. 

The Lyon group use the model atmosphere code Phoenix for application to stellar and substellar atmospheres \citep{Allard_1995, Hauschildt_1999}, which have been succesfully used to describe the evolution of low-mass stars (e.g. \citet{Baraffe_2015}). Both \citet{Chabrier_2000} and B03 presented evolutionary calculations for brown dwarfs using grids of Phoenix model atmospheres from \citet{Allard_2001}, labelled `AMES-Dusty' and `AMES-Cond', respectively. The AMES-Dusty models included dust opacity and are valid for hot (i.e. massive and/or young) brown dwarfs, whereas the AMES-Cond models neglected dust opacity representing the case where all condensates have settled below the photosphere, and are valid for cooler brown dwarfs. 

The second set of brown dwarf models we use for comparisons are from the Saumon \& Marley group, who applied and developed a 1D radiative-convective code originally designed for solar system atmospheres to brown dwarfs \citep{McKay_1989, Marley_1996, Burrows_1997, Marley_2002}. Evolutionary models from this group were presented in SM08, who varied the cloud sedimentation efficiency (see \citet{Ackerman_Marley_2001}) within their atmospheric outer boundary condition to investigate the impact of clouds on brown dwarf evolution. Here we compare our new model set to the AMES-Cond and SM08 cloud-free models, both of which  take the approach used in this work whereby condensate species are included in the chemical equilibrium calculations, but cloud opacity is neglected in the radiative transfer, modelling the scenario where dust grains settle or sediment below the photosphere.

\begin{figure}[t!]
\centering
\resizebox{1.0\hsize}{!}{\includegraphics{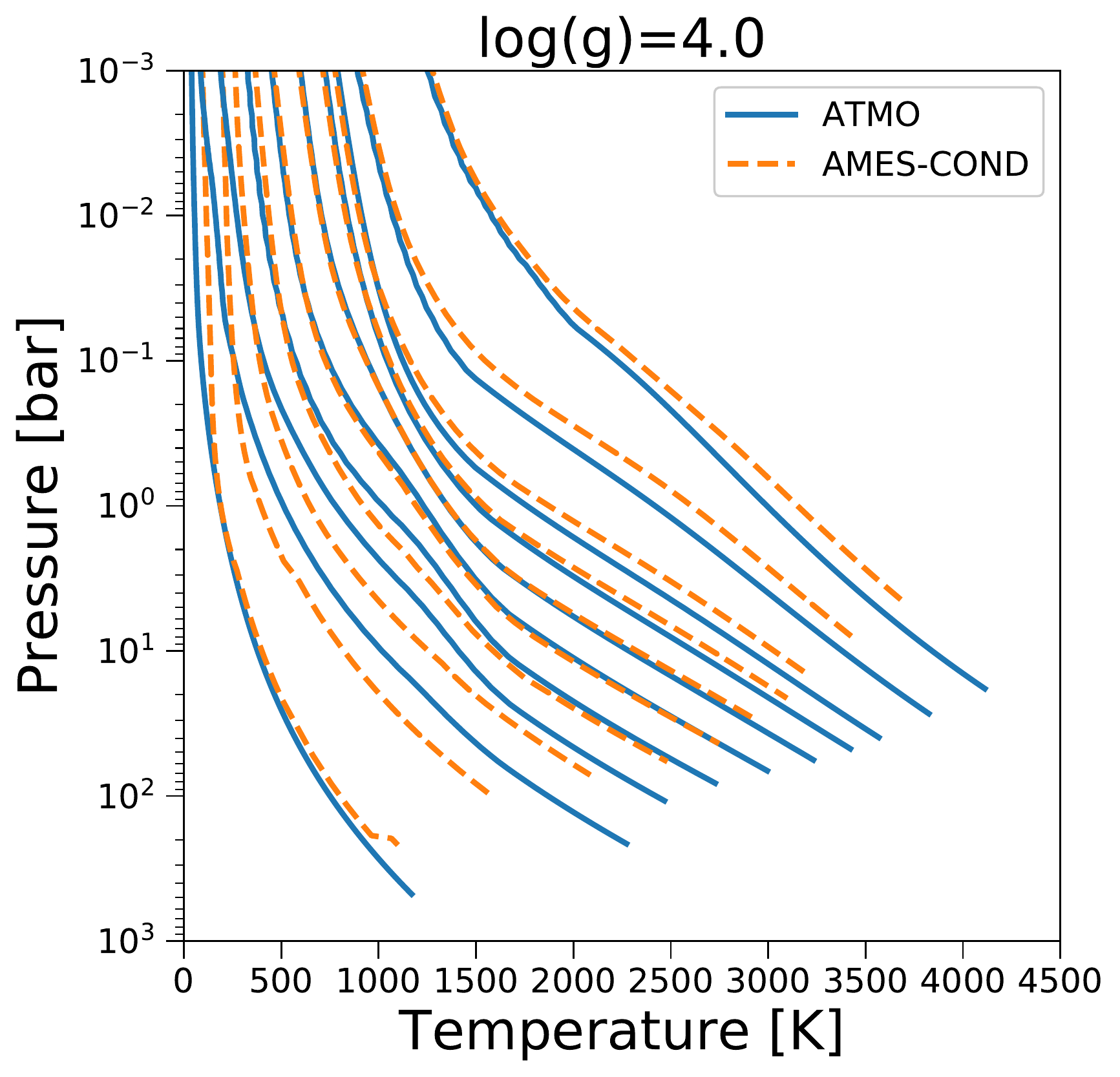}}
\caption{Self-consistent atmospheric P-T structures from this work (solid blue lines) and from the AMES-Cond models of B03 (dashed orange lines) for $\log (g)=4.0$ and $T_{\mathrm{eff}}=200$, 400, 600, 800, 1000, 1200, 1400, 1600, 2000, and 2400K (from left to right).  \label{fig:PT_profiles}}
\end{figure}


The \texttt{ATMO} and AMES-Cond atmospheric temperature profiles are compared in Figure \ref{fig:PT_profiles}, for a constant $\log(g)=4.0$ and $T_{\mathrm{eff}}$ between $200$ and $2400\,\mathrm{K}$. There are significant differences in the temperatures obtained between the models for a given $T_{\mathrm{eff}}$ and $\log(g)$, with the \texttt{ATMO} profiles being typically warmer for $T_{\mathrm{eff}}<1200\,\mathrm{K}$ and cooler for $T_{\mathrm{eff}}>1200\,\mathrm{K}$. There have been numerous model improvements that could contribute to these differences since the AMES-Cond grid was generated. Most notably improved high-temperature line lists including significantly more lines for crucial species such as $\mathrm{H_2O}$, $\mathrm{CH_4}$, and $\mathrm{NH_3}$, have increased the atmospheric opacity leading to warmer temperature profiles for $T_{\mathrm{eff}}<1200\,\mathrm{K}$. The \texttt{ATMO} $T_{\mathrm{eff}}=200\,\mathrm{K}$ model is slightly cooler in the deep atmosphere. This is likely due to the improved treatment of low-temperature equilibrium chemistry and condensation in \texttt{ATMO}, over that used in the Phoenix model at the time of generation of the AMES-Cond grid. For $T_{\mathrm{eff}}>1200\,\mathrm{K}$ the cooler \texttt{ATMO} profiles suggest we may be missing opacity at higher temperatures. We do not include the opacity of some metal oxides and metal hydrides which can be important in shaping the temperature profiles at high $T_{\mathrm{eff}}$ \citep{Malik_2019}. This is only important for high-$T_{\mathrm{eff}}$ objects (i.e. massive and/or young brown dwarfs), and will therefore not affect the evolutionary calculations of cool T--Y objects  presented in this work (see further discussion in Sections \ref{sec:grid} and \ref{sec:discussion}).

\begin{figure*}[t!]
\centering
\resizebox{0.94\hsize}{!}{\includegraphics{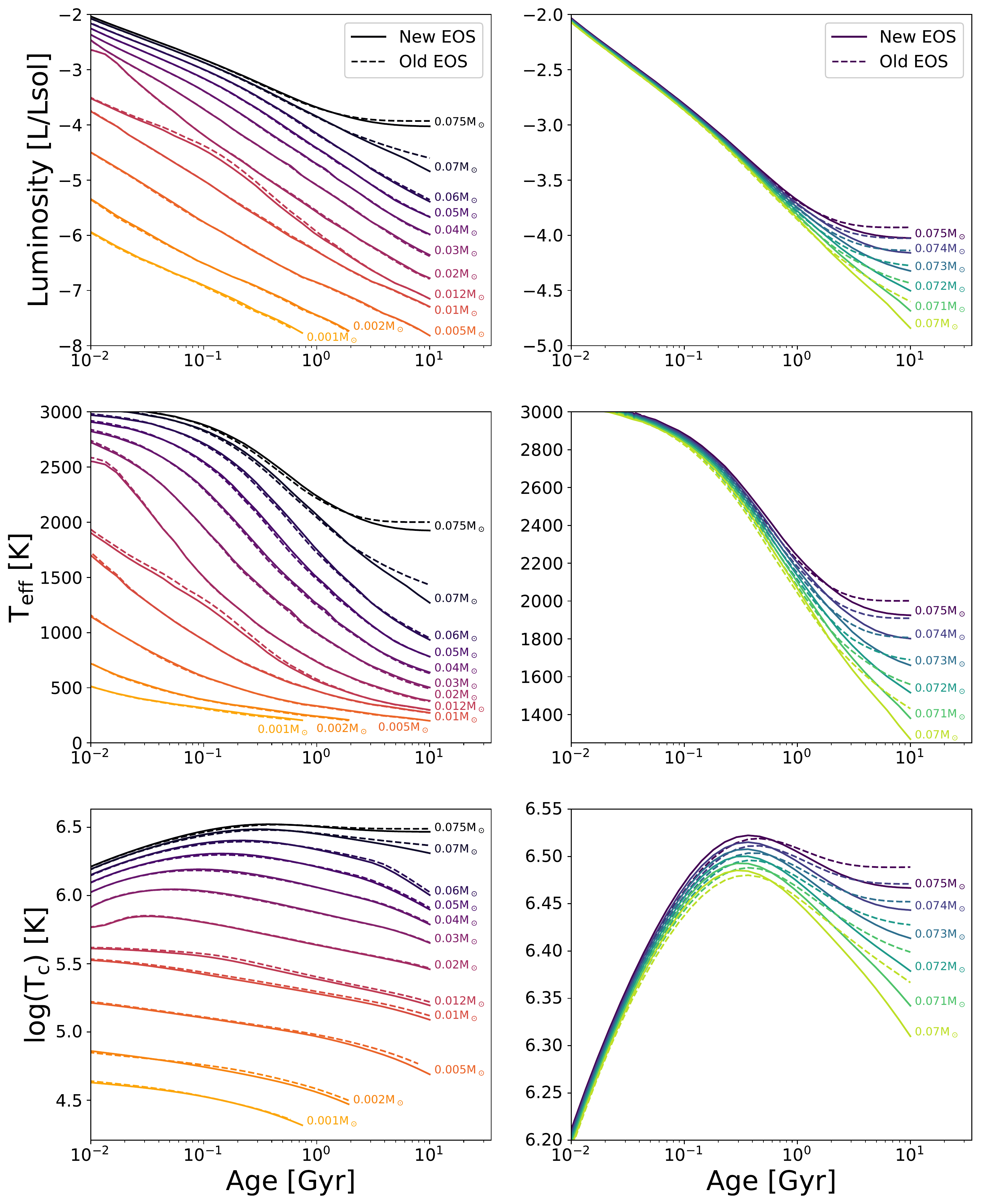}}
\caption{Luminosity, effective, and central temperature (from top to bottom) as a function of age calculated with the new EOS from \citet{Chabrier_2019} (solid lines) and the older EOS of \citet{Saumon_1995} (dashed lines). These models are calculated with \texttt{ATMO} surface boundary conditions, as described in the text. Each column displays a different selection of masses, indicated by colour-coded annotations placed next to the curves calculated with the new EOS. \label{fig:eos}}
\end{figure*}

The atmospheric temperature structures from the \texttt{ATMO} model grid are used to couple the non-grey atmosphere to the interior structure, and calculate evolutionary tracks for a range of substellar masses. One of the major improvements of the interior structure model in this work is the use of the EOS of \citet{Chabrier_2019}, over the older EOS of \citet{Saumon_1995} (see Section \ref{sec:interior_structure}). Figure \ref{fig:eos} shows evolutionary tracks calculated with these different EOSs. There are notable differences for the highest masses, with the new EOS predicting slightly cooler, less luminous objects at old ages close to the stellar--substellar transition. The new EOS also slightly changes the cooling curve around the deuterium burning minimum mass, which can be seen in the $\mathrm{0.012\,M_{\odot}}$ track.

The right  column of Figure \ref{fig:eos} shows evolutionary tracks zoomed in for objects close to the substellar boundary. The largest difference occurs for a $10\,\mathrm{Gyr}$ old $\mathrm{0.071\,M_{\odot}}$ object, which is now predicted to be $\mathrm{\sim180\,K}$ cooler in effective temperature and $\mathrm{\sim0.25\,dex}$ less luminous with the new EOS. We note, however, that we do not expect our evolutionary tracks to be accurate at the $0.001\,M_\odot$ level, as other uncertainties in the evolution model such as small changes in the helium mass fraction can cause changes to the cooling curves comparable to those caused by the new EOS. Therefore, distinguishing between the new and the old EOS will be challenging, and for this reason we avoid providing an exact value for the mass at the substellar boundary predicted by our new models.

\begin{figure}[t!]
\centering
\resizebox{\hsize}{!}{\includegraphics{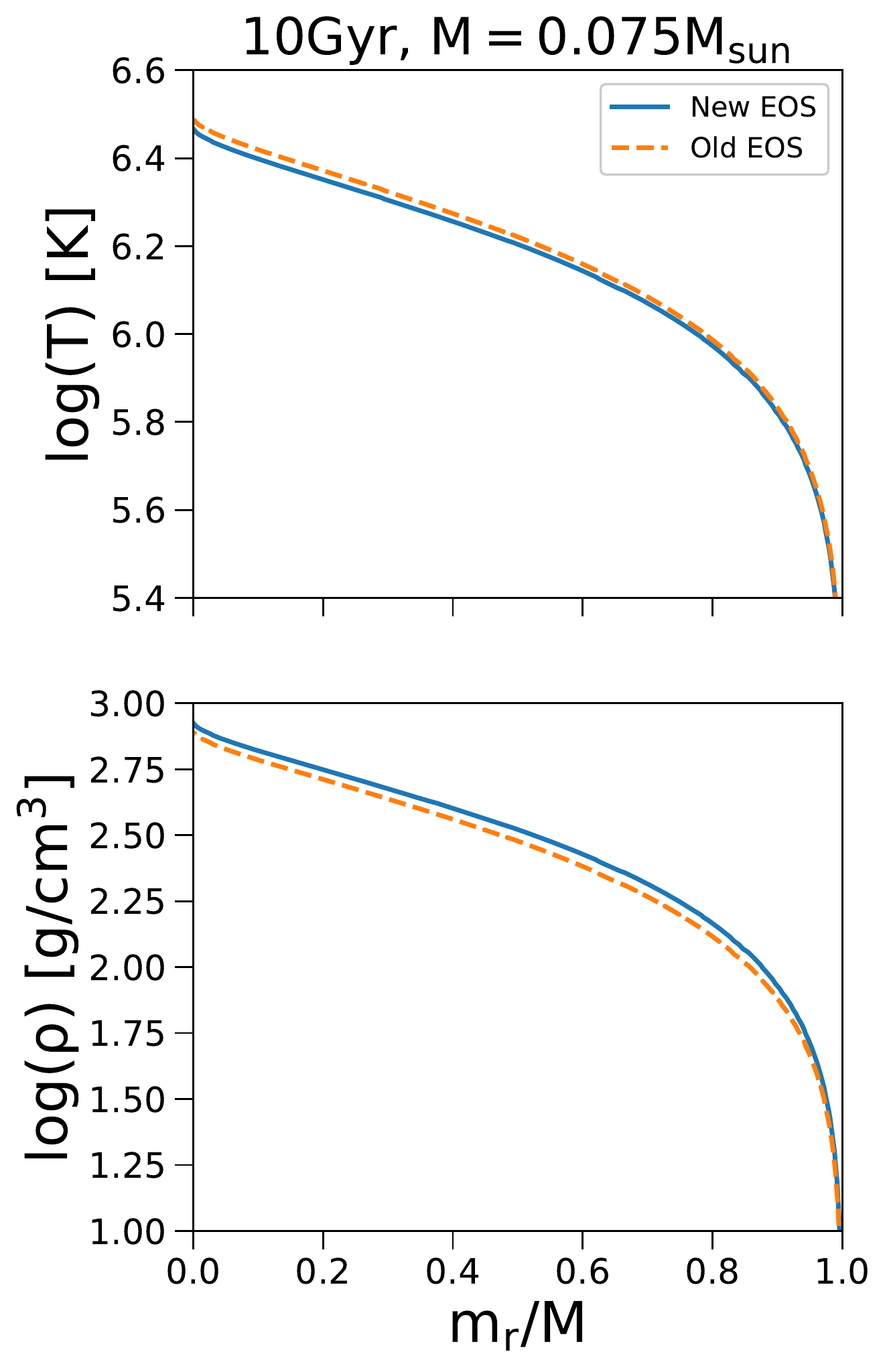}}
\caption{Interior temperature (top) and density (bottom) as a function of normalised radial mass profile of the simulated object. Solid blue lines are generated using the new EOS of \citet{Chabrier_2019}, and the dashed orange lines are generated using the older EOS of \citet{Saumon_1995}. \label{fig:interior}}
\end{figure}

 To illustrate the impact of the new EOS we show the interior temperature and density profiles of a $\mathrm{0.075\,M_{\odot}}$, $10\,$Gyr object in Figure \ref{fig:interior}. The new EOS of \citet{Chabrier_2019}  gives an object up to $\sim5\%$ cooler and $\sim8\%$ denser in the core. This therefore raises the theoretical stellar--substellar boundary by $1$-$2\%$ in mass, as the interior is now cooler and denser, thus more degenerate ($\psi\propto T/\rho^{2/3}$, where $\psi$ is the degeneracy parameter \citep{Chabrier_Baraffe_1997}) for a given mass and age. This results in a change in the cooling curves at masses near  the stellar--substellar boundary, with objects cooling to lower $T_{\mathrm{eff}}$.

\begin{figure*}[t!]
\centering
\resizebox{\hsize}{!}{\includegraphics{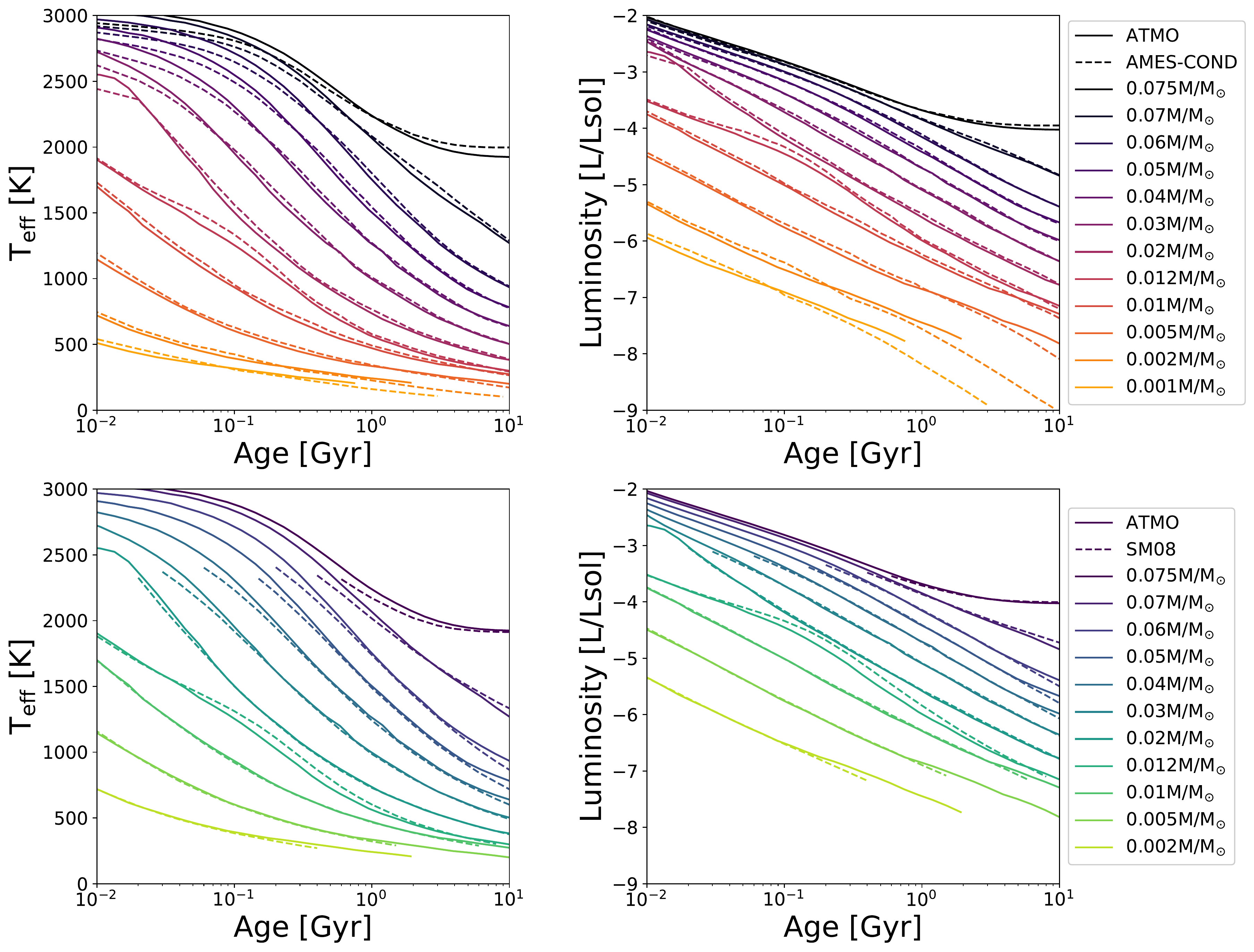}}
\caption{Evolution of the effective temperature and luminosity for a range of substellar masses from this work (solid lines), from the AMES-Cond models of B03 (dashed lines, top), and the SM08 models (dashed lines, bottom).\label{fig:evol}}
\end{figure*}

The evolutionary tracks from this work and from the AMES-Cond and SM08 calculations are compared in Figure \ref{fig:evol}, which shows the evolution of the effective temperature and luminosity for masses between 0.001 and 0.075$\,\mathrm{M_{\odot}}$. As previously discussed, the new EOS used in this work raises the hydrogen and deuterium minimum burning mass, causing changes in the cooling curves at high masses and around $0.012\,\mathrm{M_{\odot}}$, respectively. Furthermore, the shape of the evolutionary tracks of the lowest masses have changed due to the differences in the atmospheric temperature structures used as the outer boundary condition. The warmer \texttt{ATMO} temperature structures for $T_{\mathrm{eff}}<1200\,\mathrm{K}$ (Figure \ref{fig:PT_profiles}) lead to a slightly cooler, less luminous $0.001\,\mathrm{M_{\odot}}$ object for ages $\mathrm{<0.1\,Gyr}$, and a warmer, brighter object for ages $\mathrm{>0.1\,Gyr}$. We find qualitatively similar differences when comparing \texttt{ATMO} to the AMES-Cond and to the SM08 tracks, respectively.

\subsection{Comparison with dynamical masses} \label{sec:dyn_masses}

Dynamical mass measurements of brown dwarfs from astrometric monitoring programs of binary systems provide useful tests for evolutionary models (e.g. \citet{Dupuy_liu_2017}). Recently, \citet{Brandt_2019} presented a dynamical mass measurement of the first imaged brown dwarf Gl 229 B of $70\pm5\,\mathrm{M_{Jup}}$. This measurement joins a growing list of massive T dwarfs that are challenging evolutionary models (e.g. \citet{Bowler_2018, Dieterich_2018, Dupuy_2019}). We note, however, that the dynamical mass measurement of Gl 229 B should be considered with caution until confirmed unambiguously (the paper is only on ArXiv for now) (R. Oppenheimer priv. comm). 

\begin{figure}[t!]
\centering
\resizebox{1.0\hsize}{!}{\includegraphics{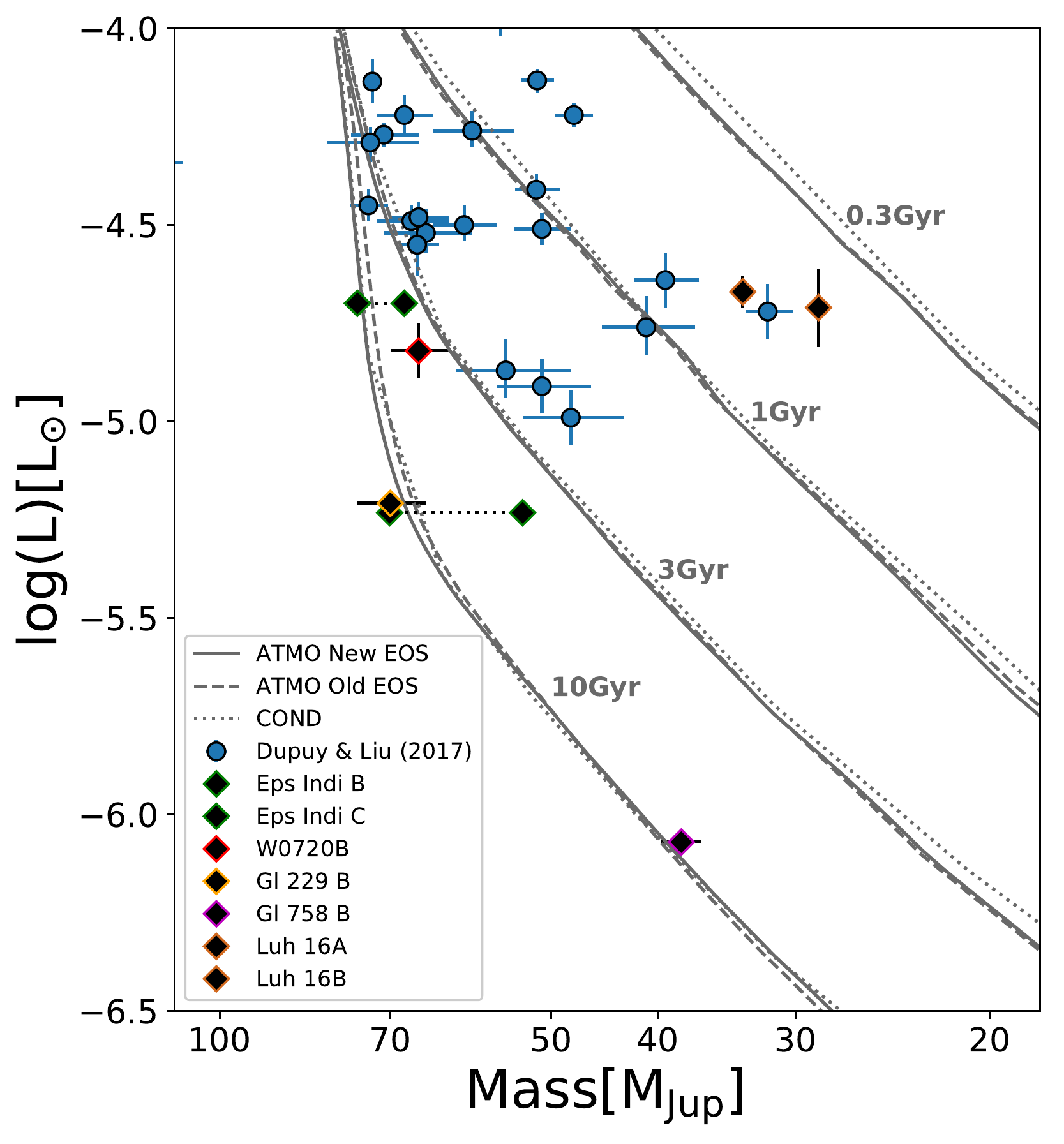}}
\caption{Luminosity as a function of mass for ultracool dwarfs that have dynamical mass measurements. \texttt{ATMO} isochrones from this work calculated using the new EOS from \citet{Chabrier_2019} and the older EOS of \citet{Saumon_1995} are plotted as solid and dashed grey lines respectively. Isochrones from the AMES-Cond models of B03 are plotted as dotted grey lines. Most mass measurements are from \citet{Dupuy_liu_2017} and are plotted as blue circles, with other literature measurements plotted as black diamonds \citep{Bowler_2018, Lazarenko_2018, Dupuy_2019, Brandt_2019}. Both mass measurements of $\epsilon$ Indi BC from \citet{Cardoso_2012} and \citet{Dieterich_2018} are plotted, and we refer the reader to \citet{Dupuy_2019} for a discussion on these conflicting mass measurements. We indicate key objects with coloured outlines. \label{fig:dyn_mass}}
\end{figure}

We show in Figure \ref{fig:dyn_mass} the luminosity as a function of mass for ultracool dwarfs with dynamical mass measurements including Gl 229 B. In this figure we show isochrones from this work calculated with the new and old EOSs, and isochrones from B03. The new EOS can be seen predicting cooler, less luminous objects in this figure for old high-mass objects. For a $\mathrm{70\,M_{Jup}}$ object at an age of $\mathrm{10\,Gyr}$, the \texttt{ATMO} tracks calculated with the new EOS are $\sim0.1\,\mathrm{dex}$ less luminous than the AMES-Cond tracks of B03, and $\sim0.4\,\mathrm{dex}$ less luminous than the hybrid cloud tracks of SM08.

As discussed by \citet{Brandt_2019}, the evolutionary models of B03 and SM08 are only compatible with a mass of $70\pm5\,\mathrm{M_{Jup}}$ for Gl 229 B if the system is old (7-10\,Gyr), in some tension with the $2$-$6\,$Gyr age estimate of the host star from kinematics and stellar activity. The \texttt{ATMO} tracks calculated with the new EOS may help relieve some of the tension surrounding the age of the system given that high-mass objects are predicted to be cooler and less luminous at a given age. We note, however,  that the difference between the old and new EOS is not observationally significant given the uncertainty on the mass measurement of Gl 229 B shown in Figure \ref{fig:dyn_mass}.


\subsection{Equilibrium and non-equilibrium chemistry models}\label{sec:NEQ_spec}

 Non-equilibrium processes primarily affect the carbon and nitrogen chemistry of the atmosphere of a cool T--Y-type object  \citep{Zahnle_Marley_2014}. Mixing processes can be responsible for bringing CO and $\mathrm{N_2}$ into the upper atmosphere where $\mathrm{CH_4}$ and $\mathrm{NH_3}$ should be the dominant carbon- and nitrogen-bearing species according to thermodynamic equilibrium. The chemical timescale to convert $\mathrm{CO}\to\mathrm{CH_4}$ and $\mathrm{N_2}\to\mathrm{NH_3}$ is typically long compared to mixing timescales, meaning excess CO is predicted in T-type objects \citep{Fegley_Lodders_1996} and depleted ammonia in cooler late T--Y-type objects  \citep{Zahnle_Marley_2014, Tremblin_2015}. Indeed, observational studies have revealed an excess of CO in both the atmosphere of Jupiter \citep{Bezard_2002} and T dwarfs \citep{Noll_1997, Geballe_2009}, and depleted $\mathrm{NH_3}$ in late T and Y dwarfs \citep{Saumon_2000, Saumon_2006, Leggett_2015, Tremblin_2015}.
 
\begin{figure*}[h!]
\centering
\resizebox{\hsize}{!}{\includegraphics{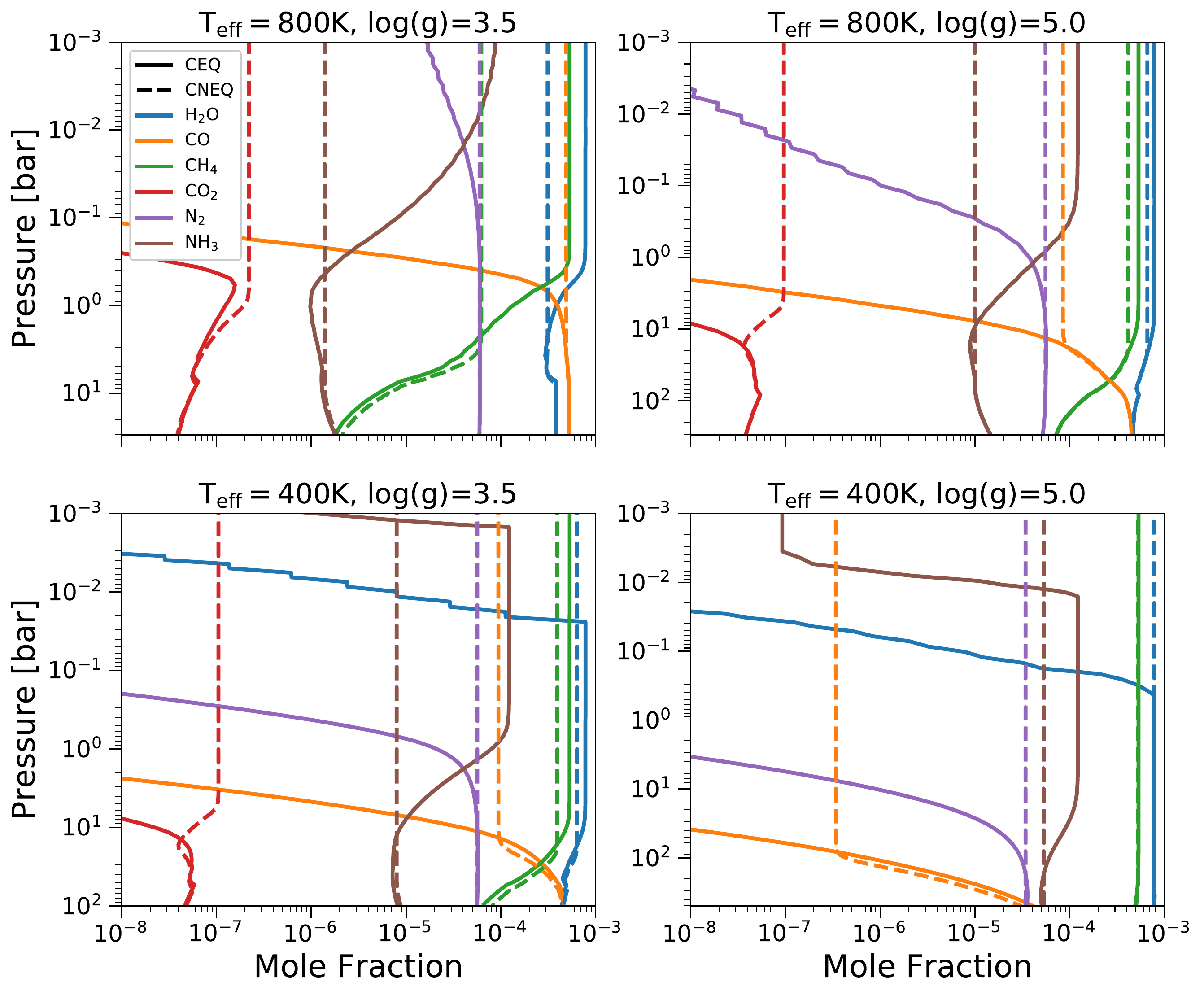}}
\caption{Chemical abundance profiles of $\mathrm{H_2O}$, $\mathrm{CO}$, $\mathrm{CH_4}$, $\mathrm{CO_2}$, $\mathrm{N_2}$, and $\mathrm{NH_3}$ of self-consistent \texttt{ATMO} models generated under the assumption of chemical equilibrium (solid lines) and non-equilibrium chemistry due to vertical mixing (dashed lines). Non-equilibrium models are calculated with the strong $K_{\mathrm{zz}}$ mixing relationship with surface gravity, as shown in Figure \ref{fig:Kzz_plot}. The rows display models with different effective temperatures, and the columns display models with different surface gravities, as indicated in the plot titles. In the top left $T_{\mathrm{eff}}=800\,\mathrm{K}$, $\log(g)=3.5$ plot, the non-equilibrium abundance of $\mathrm{CH_4}$ lies below the $\mathrm{N_2}$ abundance. \label{fig:NEQ_chemistry}}
\end{figure*}

\begin{figure*}[h!]
\centering
\resizebox{\hsize}{!}{\includegraphics{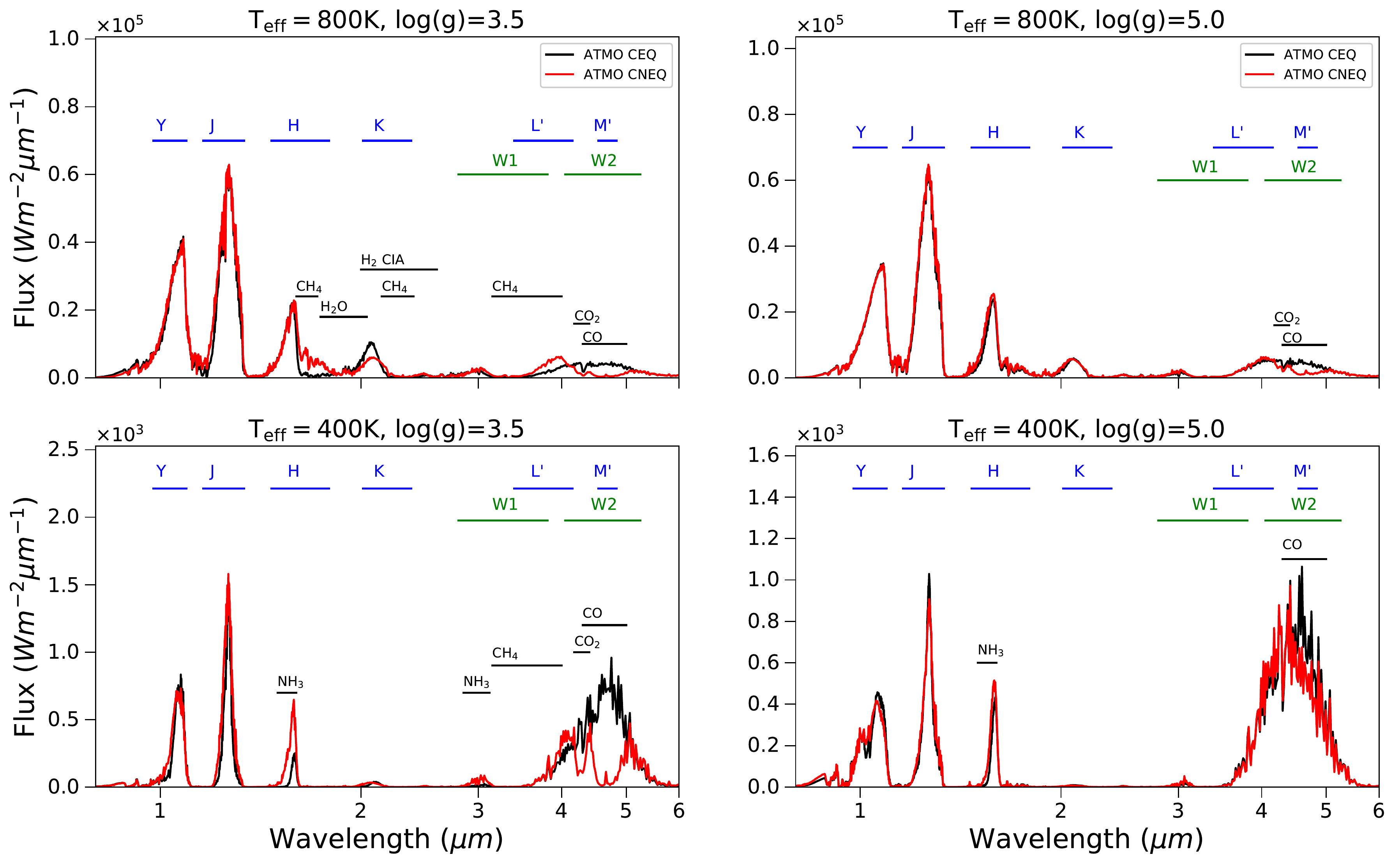}}
\caption{Emission spectra of \texttt{ATMO} model atmospheres generated under the assumption of chemical equilibrium (black) and non-equilibrium chemistry due to vertical mixing (red). Non-equilibrium models are calculated with the strong $K_{\mathrm{zz}}$ mixing relationship with surface gravity, as shown in Figure \ref{fig:Kzz_plot}. The rows display models with different effective temperatures, and the columns display models with different surface gravities, as indicated in the plot titles. Overplotted for clarity are the approximate locations of molecular absorption features causing differences between the equilibrium and non-equilibrium spectra. Also indicated in the plots are the locations of the Mauna Kea near-infrared photometric filters (blue bars), and the WISE infrared filters (green bars).\label{fig:NEQ_spectra}}
\end{figure*}

 Figures \ref{fig:NEQ_chemistry} and \ref{fig:NEQ_spectra} show chemical abundance profiles and synthetic emission spectra, respectively, calculated assuming chemical equilibrium (CEQ) and consistent non-equilibrium chemistry (CNEQ) due to vertical mixing for a sample of effective temperatures and surface gravities. We
note that, as discussed in Section \ref{sec:chem}, we scale the eddy diffusion coefficient $K_{\mathrm{zz}}$ with surface gravity such that vertical mixing is stronger in lower gravity objects. As such the differences in the spectra presented in Figure \ref{fig:NEQ_spectra} are larger for lower gravity models. Furthermore, in these figures we show non-equilibrium models calculated with the strong $K_{\mathrm{zz}}$ mixing relationship (Figure \ref{fig:Kzz_plot}), to maximise the differences between the equilibrium and non-equilibrium spectra.
 
The abundance of $\mathrm{CH_4}$ is quenched by approximately an order of magnitude in the upper atmosphere for the $T_{\mathrm{eff}}=800\,\mathrm{K}$, $\log(g)=3.5$ model in Figure \ref{fig:NEQ_chemistry}. The depleted $\mathrm{CH_4}$ abundance lowers the opacity in the absorption bands at $\sim1.6\,\mu m$ and $\sim3.15\,\mu m$ giving brighter $H$  and $L^\prime$ bands in the non-equilibrium spectrum of this model in Figure \ref{fig:NEQ_spectra}. The $K$-band flux is lower in the non-equilibrium spectrum due to the P-T profile being $\sim150\,\mathrm{K}$ cooler at 1\,bar than the equilibrium model. This causes the model levels in which the $K$-band flux is generated to be shifted to slightly higher pressures where the $\mathrm{H_2-H_2}$ collisionally induced absorption is stronger. The weaker mixing in the high-gravity $T_{\mathrm{eff}}=800\,\mathrm{K}$, $\log(g)=5.0$ model means that $\mathrm{CH_4}$ is not depleted as strongly (Figure \ref{fig:NEQ_chemistry}). This smaller change in $\mathrm{CH_4}$ abundance combined with the near-infrared photosphere lying deeper in the atmosphere for higher gravity means that there is no change in the spectrum in the $H$, $K$, and $L^\prime$ bands, as there was in the lower gravity $T_{\mathrm{eff}}=800\,\mathrm{K}$, $\log(g)=3.5$ model.

The abundances of $\mathrm{CO}$ and $\mathrm{CO_2}$ are increased by many orders of magnitude in the upper atmosphere under non-equilibrium chemistry due to vertical mixing in all models shown in Figure \ref{fig:NEQ_chemistry}. Despite $\mathrm{CO_2}$ being several orders of magnitude less abundant than CO, both $\mathrm{CO}$ and $\mathrm{CO_2}$ have strong absorption features at $\sim4.3\,\mu m$ and $\sim4.18\,\mu m$, respectively, and their increased abundances lower the flux at these wavelengths in the $W2$ and $M^\prime$ bands in the non-equilibrium spectra shown in Figure \ref{fig:NEQ_spectra}. The abundance of $\mathrm{NH_3}$ is quenched in the models shown in Figure \ref{fig:NEQ_chemistry} under non-equilibrium chemistry, and can be seen having an effect on the predicted spectrum of the cooler $T_{\mathrm{eff}}=400\,\mathrm{K}$ models shown in Figure \ref{fig:NEQ_spectra}. The depleted $\mathrm{NH_3}$ abundance lowers the opacity in the $\mathrm{NH_3}$ absorption bands at $\sim1.5\,\mu m$ and $\sim2.85\,\mu m$ giving a brighter $H$ band and more flux at $\sim3\,\mu m$ in the non-equilibrium spectrum. 
 
\subsection{Colour-magnitude diagrams} \label{sec:CMDs}

\begin{figure*}[h!]
\centering
\resizebox{\hsize}{!}{\includegraphics{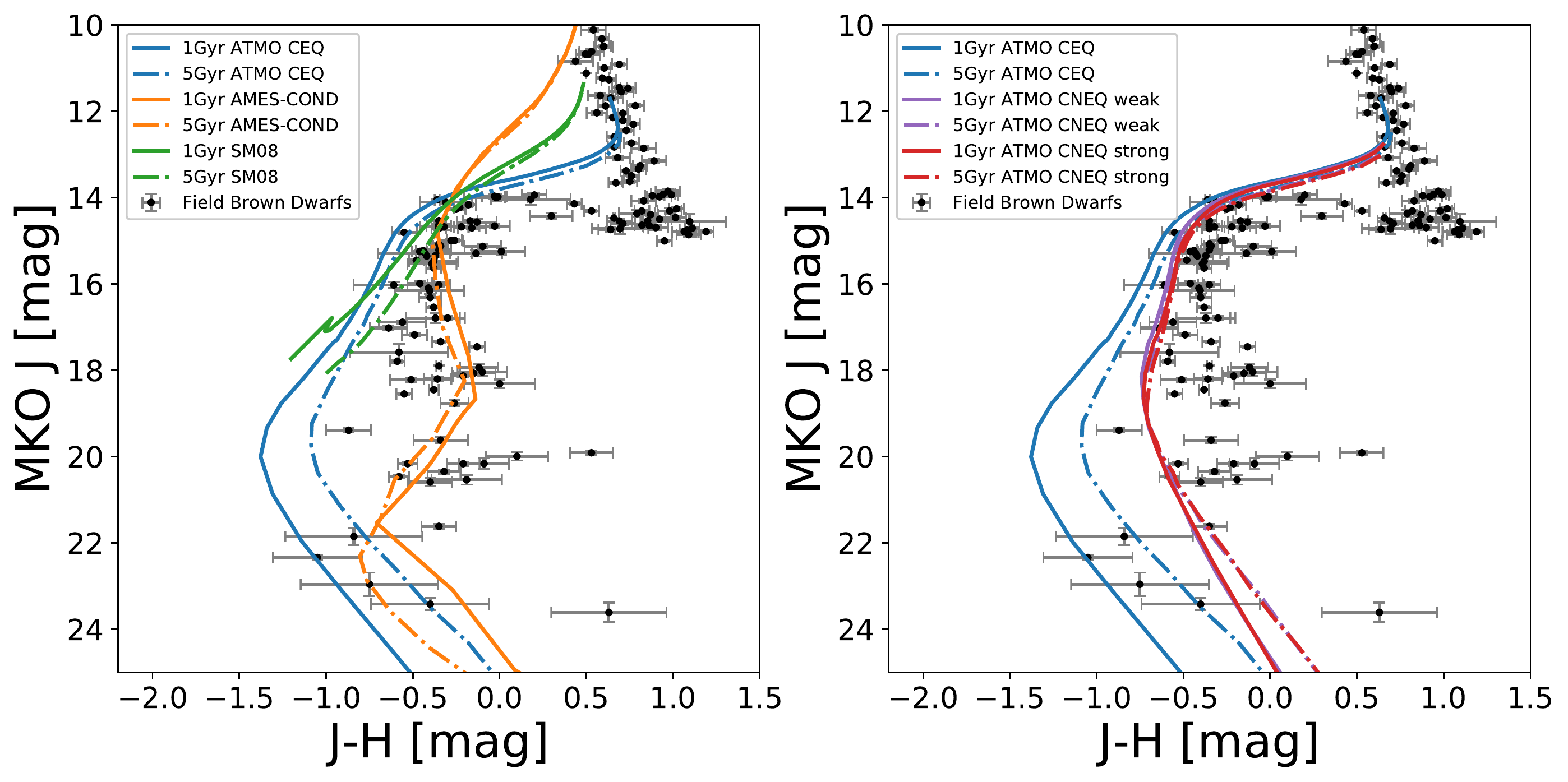}}
\caption{$J-H$ colour as a function of absolute $J$-band magnitude. Left: Isochrones from chemical equilibrium models of this work (blue), B03 (orange), and SM08 (green). Right: Isochrones from chemical equilibrium models (blue) and non-equilibrium chemistry models with weak (purple) and strong (red) vertical mixing  from this work. In both panels the photometry of field brown dwarfs are plotted as black circles, with the data taken from the database of ultracool parallaxes. The data is filtered to only include field brown dwarfs from \citet{Dupuy_Liu_2012} and \citet{Dupuy_Kraus_2013}. \label{fig:CMD_J-H}}
\end{figure*}

\begin{figure*}[h!]
\centering
\resizebox{\hsize}{!}{\includegraphics{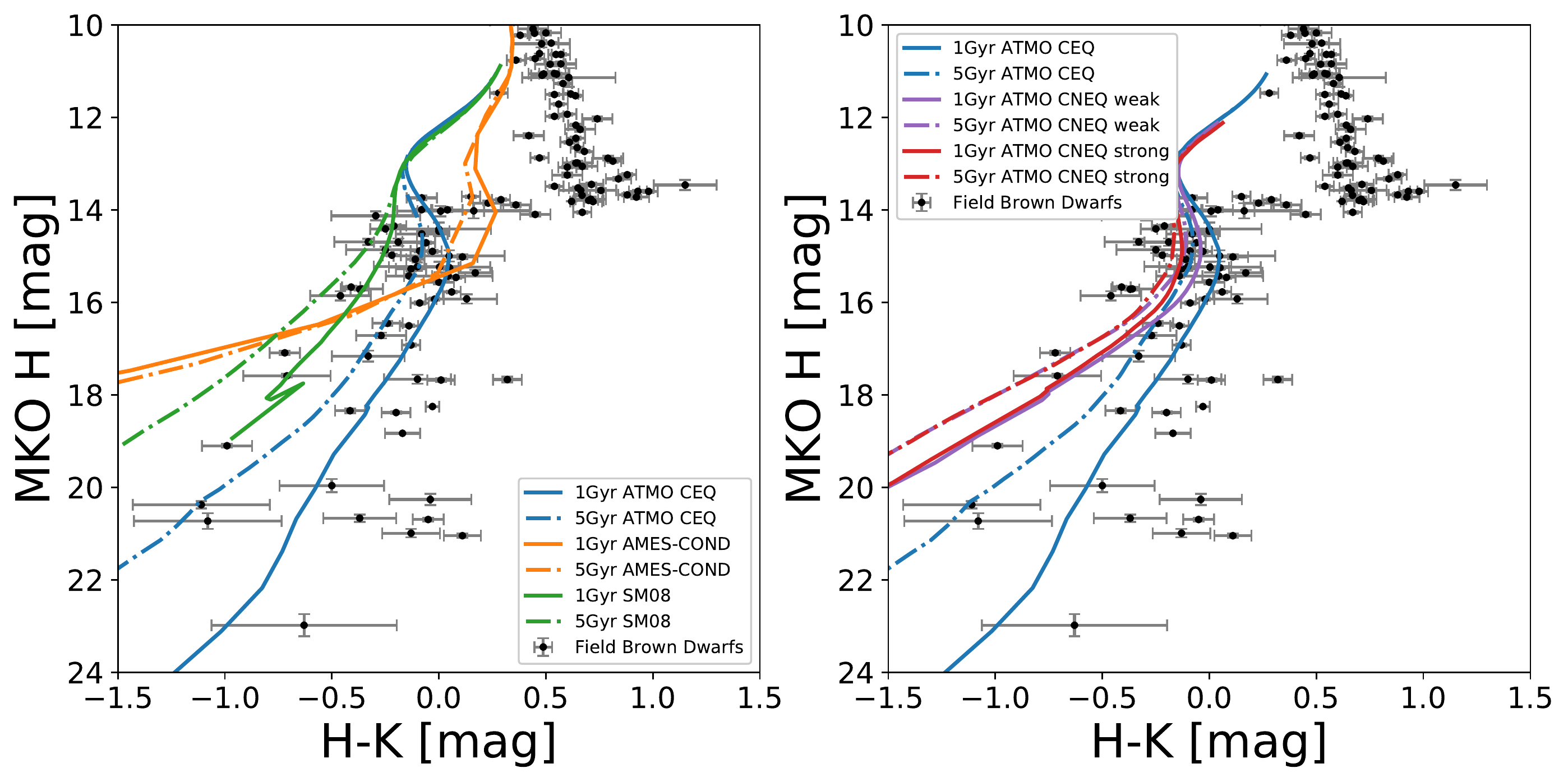}}
\caption{Same as Figure \ref{fig:CMD_J-H}, but for the $H-K$ colour as a function of absolute $H$-band magnitude. Isochrones from chemical equilibrium models are compared in the left panel, and the effect of non-equilibrium chemistry on isochrones is shown in the right panel.\label{fig:CMD_H-K}}
\end{figure*}

\begin{figure*}[h!]
\centering
\resizebox{\hsize}{!}{\includegraphics{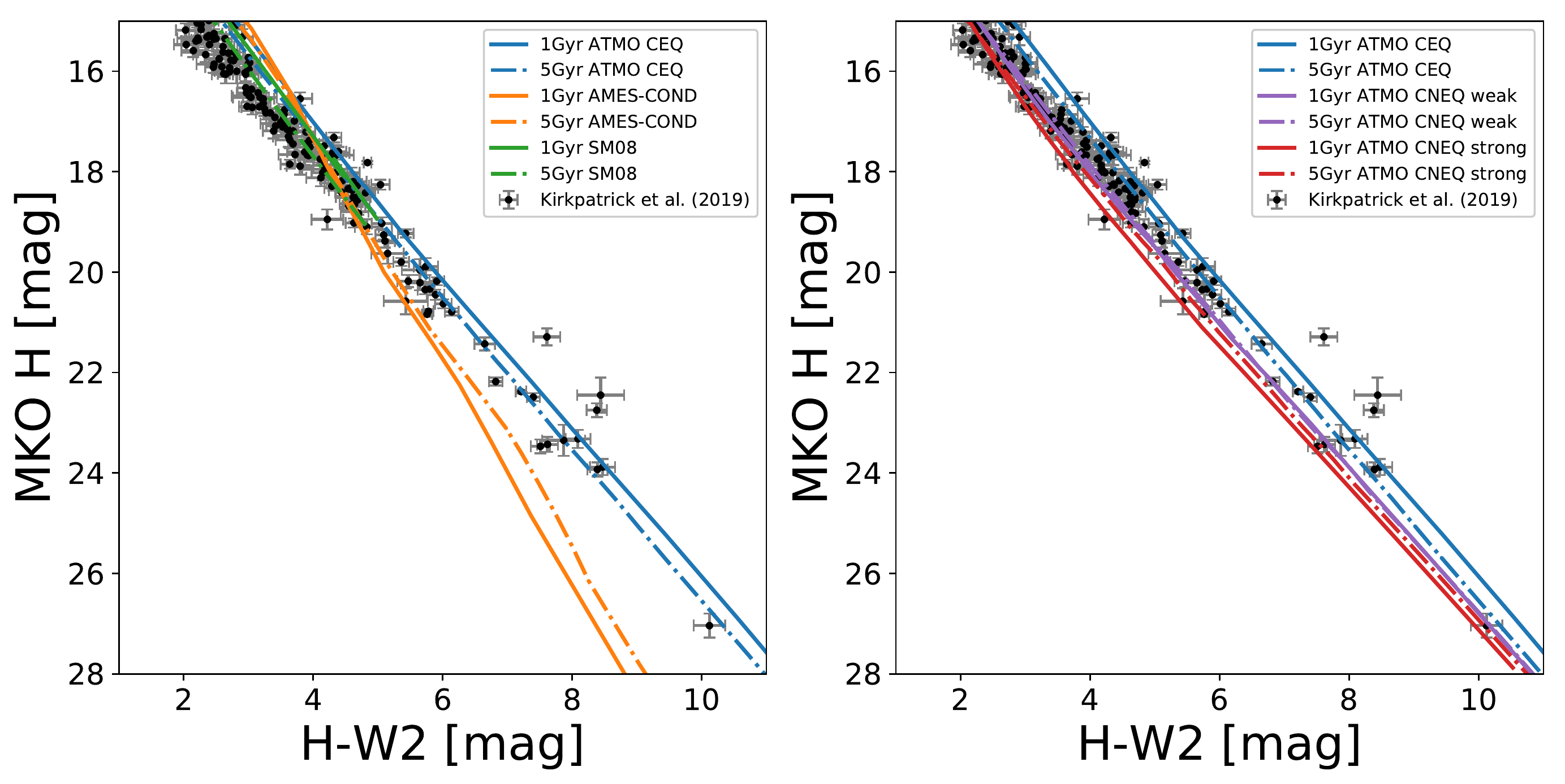}}
\caption{$H-W2$ colour as a function of absolute $H$-band magnitude, with photometry of late T and Y dwarfs within $20\,\mathrm{pc}$ from \citet{Kirkpatrick_2019} plotted as black circles. Overplotted in the left panel are isochrones of chemical equilibrium models from this work (in blue), from B03 (in orange), and from SM08 (in green). Overplotted in the right panel are isochrones of chemical equilibrium models (in blue), and non-equilibrium chemistry models with weak and strong vertical mixing (in purple and red, respectively). \label{fig:K19_cmd}}
\end{figure*}

\begin{figure*}[h!]
\centering
\resizebox{\hsize}{!}{\includegraphics{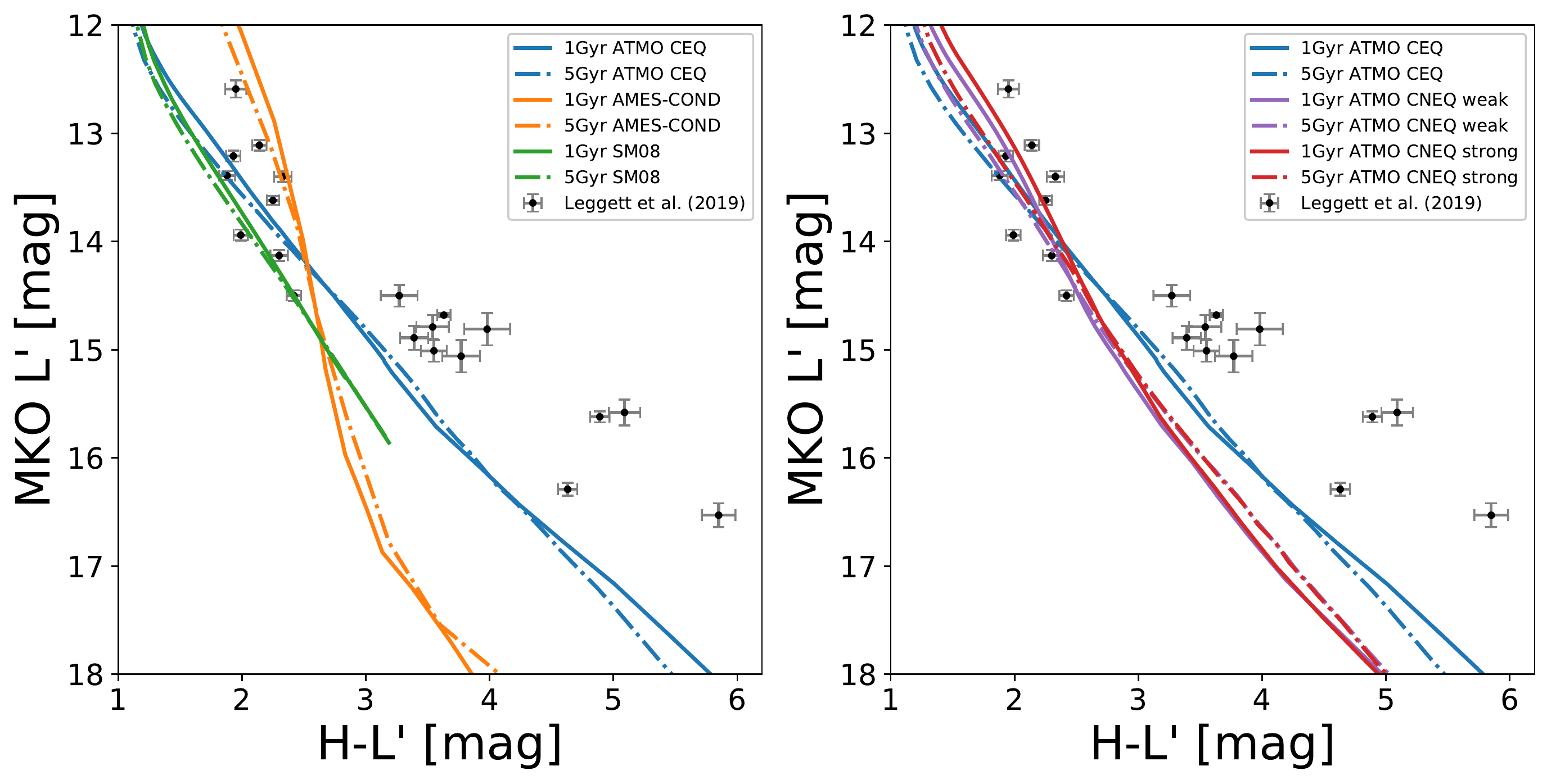}}
\caption{$H-L'$ colour as a function of absolute $L'$ band magnitude, with photometry of late T and Y dwarfs from \citet{Leggett_2019} plotted as black circles. Overplotted in the left panel are isochrones of chemical equilibrium models from this work (in blue), from B03 (in orange), and from SM08 (in green.) Overplotted in the right panel are isochrones of chemical equilibrium models (in blue), and non-equilibrium models with weak and strong vertical mixing (in purple and red, respectively). \label{fig:L19_cmd}}
\end{figure*}

In this section we compare the new model set presented in this work, along with the models of B03 and SM08, to observational datasets in colour-magnitude diagrams. In each of the colour-magnitude diagrams presented in Figures \ref{fig:CMD_J-H} -- \ref{fig:L19_cmd}, the left panel shows isochrones of photometry derived from chemical equilibrium atmosphere models from each of these three works, and the right panel shows isochrones of the equilibrium and non-equilibrium models of this work to illustrate the impact of vertical mixing on the predicted colours of cool brown dwarfs. In Figures \ref{fig:CMD_J-H} and \ref{fig:CMD_H-K} we present near-infrared colour-magnitude diagrams including photometry from the database of ultracool parallaxes \citep{Dupuy_Liu_2012, Dupuy_Kraus_2013}. We exclude from the dataset the known and suspected binaries, young low-gravity objects, and low-metallicity objects.

Figure \ref{fig:CMD_J-H} shows the $J-H$ colours as a function of absolute $J$ magnitude. The data show the M and L dwarf population for $J<14$, which gets progressively redder for cooler objects, and the sharp change to bluer colours for the methane dominated T dwarfs at $J\sim14.5$ known as the L-T transition. The cool T and Y dwarf objects for which the models presented in this work are most applicable, lie below $J\sim15$, and their $J-H$ colours are best reproduced by the AMES-Cond models, with both the \texttt{ATMO} and SM08 isochrones predicting colours that are too blue compared to the data. However, this is  caused by the outdated physics used within the AMES-Cond models, which lack $\mathrm{CH_4}$ and $\mathrm{NH_3}$ opacity in the $H$ band due to the incomplete line lists used at the time. The brighter AMES-Cond $H$ band therefore gives redder $J-H$ colours that coincidentally more closely match the data compared to the \texttt{ATMO} and SM08 tracks, which use more complete $\mathrm{CH_4}$ and $\mathrm{NH_3}$ line lists. These more complete line lists have added opacity to the $H$ band since the generation of the AMES-Cond models.

As discussed in Section \ref{sec:NEQ_spec} and shown in Figure \ref{fig:NEQ_spectra}, non-equilibrium chemistry due to vertical mixing can quench $\mathrm{CH_4}$ and $\mathrm{NH_3}$, lowering the opacity and increasing the flux through the $H$ band compared to models calculated in chemical equilibrium. As shown in the right panel of Figure \ref{fig:CMD_J-H}, vertical mixing reddens the predicted $J-H$ colours compared to chemical equilibrium tracks, moving the isochrones towards the observed colours of T dwarfs. The difference between the weak and strong vertical mixing tracks is small. This is due to the abundance of methane only varying by a small factor at the quench levels corresponding to the differing $K_{\mathrm{zz}}$ values, hence giving similar $H$ band magnitudes. Despite the non-equilibrium chemistry models improving the $J-H$ colours through $H$-band brightening, additional physics not included in this work such as reductions in the temperature gradient due to thermochemical instabilities \citep{Tremblin_2015, Tremblin_2016} and/or cloud opacity \citep{Morley_2012} can reduce the flux in the $J$ band and better reproduce the red $J-H$ colours of late T dwarfs. For cooler objects, the chemical equilibrium tracks begin to redden and reconverge with the observed colours of the Y dwarfs, which lie below $J\sim21$ in Figure \ref{fig:CMD_J-H}.

Figure \ref{fig:CMD_H-K} shows the $H-K$ colours as a function of absolute $H$-band magnitude. Similarly to Figure \ref{fig:CMD_J-H}, the reddening M-L sequence along with the sharp L-T transition to bluer colours is shown by the data. The cool T--Y-type  objects lying below $H\sim15$ have $H-K$ colours best reproduced by the \texttt{ATMO} isochrones, with the AMES-Cond and SM08 tracks predicting colours that are too blue compared to the data. In addition to the differing $\mathrm{CH_4}$ and $\mathrm{NH_3}$ opacity and line lists used in the models causing differences in the $H$ band, the updated $H_2-H_2$ collisionally induced absorption used in the \texttt{ATMO} models is responsible for altering the $K$-band magnitude and improving the comparison with the observations in this diagram (see also \citet{Saumon_2012}). Unlike the $J-H$ colours in Figure \ref{fig:CMD_J-H}, including non-equilibrium chemistry due to vertical mixing moves the isochrones away from the observed $H-K$ colours. This is due to the quenching of $\mathrm{CH_4}$ and $\mathrm{NH_3}$ brightening the $H$ band, as shown in Figure \ref{fig:NEQ_spectra}. Again, the difference between the weak and strong vertical mixing tracks is small, due to the reasons mentioned above. Similarly to the $J-H$ colours, the $H-K$ colours could be improved by temperature gradient reductions and/or cloud opacity, neither of which is included in this work.

As noted by \citet{Leggett_2019}, cool T--Y-type brown dwarfs emit a large percentage of their total energy through the $3.5$ to $5.5\,\mu m$ flux window, at longer wavelengths probed by the $J$, $H$, and $K$ filters considered in Figures \ref{fig:CMD_J-H} and \ref{fig:CMD_H-K}. The \textit{WISE} $W1$ and $W2$ filters probe this wavelength region and can provide useful photometry by which to characterise cool brown dwarfs. We show the $H$ versus $H-W2$ colour-magnitude diagram in Figure \ref{fig:K19_cmd}, which contains data points from \citet{Kirkpatrick_2019}. \citet{Kirkpatrick_2019} presented new measurements of trigonometric parallaxes of late T and Y dwarfs with the \textit{Spitzer} space telescope, and combined these measurements with others published in the literature to complete a sample of $\geq\mathrm{T6}$ dwarfs within $20\,\mathrm{pc}$. Within this sample \citet{Kirkpatrick_2019} found a tight correlation of the $H-W2$ colour with absolute $H$-band magnitude (see their figure 8), which provides a useful metric for benchmarking atmosphere models. 

The $H-W2$ colours in Figure \ref{fig:K19_cmd} are nicely reproduced by the \texttt{ATMO} isochrones in comparison to the AMES-Cond tracks, which become too blue for objects fainter than $\sim19$ magnitude due to missing opacity in the $H$ band from the incomplete $\mathrm{CH_4}$ and $\mathrm{NH_3}$ line lists used. The \texttt{ATMO} chemical equilibrium tracks, even though  a great improvement over the AMES-Cond models, are slightly too red compared to the data. Including non-equilibrium chemistry due to vertical mixing reduces the $W2$ band magnitude due to increased CO and $\mathrm{CO_2}$ absorption and increases the $H$-band magnitude due to the quenching of $\mathrm{CH_4}$ and $\mathrm{NH_3}$ (see Figure \ref{fig:NEQ_spectra}), resulting in bluer $H-W2$ colours. The weak vertical mixing tracks provide the best comparison to the data;  the strong mixing tracks are slightly too blue compared to the data. This indicates that the $H-W2$ colours of brown dwarfs can be used to calibrate vertical mixing and constrain the values of $K_{\mathrm{zz}}$ that should be used in atmosphere models.

A long-standing issue in the understanding of cool brown dwarfs is known as the 4 micron problem, whereby the $\lambda\sim4\,\mu m$ model fluxes are too low compared to observations of cool brown dwarfs \citep{Leggett_2012, Leggett_2013, Leggett_2015, Leggett_2017}. This problem has most recently been demonstrated and discussed by \citet{Leggett_2019}, who presented new $L^\prime$ photometry of a sample of late T and Y dwarfs showing that the $\lambda\sim4\mu m$ discrepency occurs in objects cooler than $T_{\mathrm{eff}}\sim700\,\mathrm{K}$ and increases towards lower $T_{\mathrm{eff}}$. We show in Figure \ref{fig:L19_cmd} the $H-L'$ colours as a function of absolute $L'$ band magnitude with photometric data points from \citet{Leggett_2019}. All the chemical equilibrium models in the left panel of Figure \ref{fig:L19_cmd} underpredict the $L^\prime$ magnitude for  $H-L^\prime>3$, corresponding to objects with $T_{\mathrm{eff}}<600\,\mathrm{K}$ using the \texttt{ATMO} $1\,\mathrm{Gyr}$ isochrone. Including non-equilibrium chemistry due to vertical mixing can increase the $\lambda\sim4\,\mu m$ flux as $\mathrm{CH_4}$ is quenched in the atmosphere, lowering the opacity at this wavelength (see Figure \ref{fig:NEQ_spectra}). However, as the $H$ band also brightens when including vertical mixing the $H-L^\prime$ colours become bluer, moving the tracks away from the observed population of late T and Y dwarfs compared to the chemical equilibrium tracks. As noted by \citet{Leggett_2019} and \citet{Morley_2018}, the discrepancy between the models and the observed $\lambda\sim4\,\mu m$ flux is likely due to processes happening in these atmospheres that are not currently captured by 1D radiative-convective models, such as thermochemical instabilities, cloud clearing, or breaking gravity waves.

\subsection{Spectral comparisons with other models}\label{sec:spectra}

\begin{figure}[t!]
\centering
\resizebox{1.0\hsize}{!}{\includegraphics{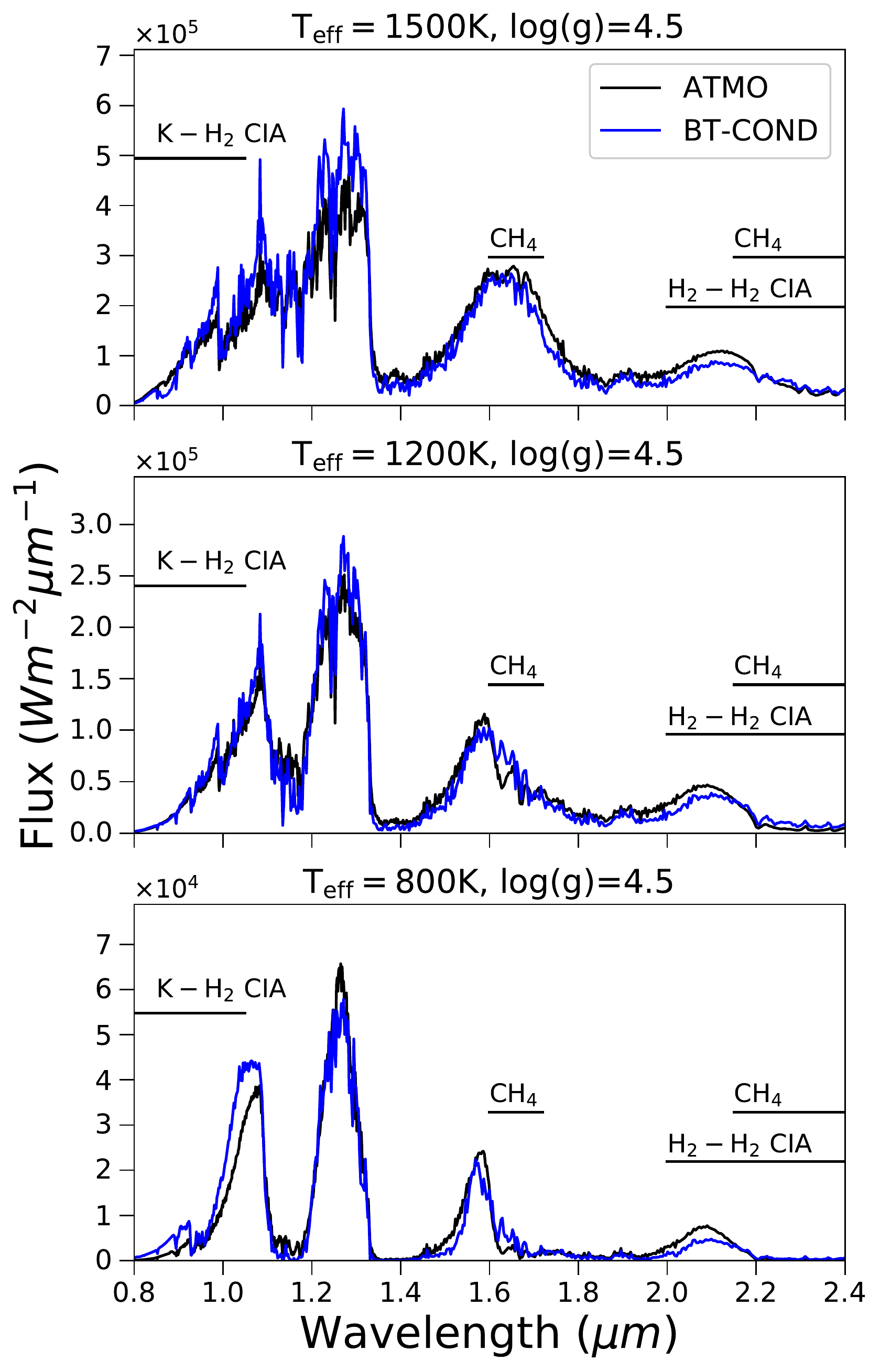}}
\caption{Synthetic near-infrared emission spectra from \texttt{ATMO} compared with models from the BT-Cond grid \citep{Allard_2012} for a range of effective temperatures.\label{fig:BT-COND_spec}}
\end{figure}

The AMES-Cond grid of the Lyon group  was labelled as such due to the NASA-AMES line lists used to calculate the opacity of $\mathrm{H_2O}$ and TiO. Since the calculation of these models there have been significant improvements in high-temperature line lists for these species, in particular the BT2 $\mathrm{H_2O}$ line list from \citet{Barber_2006}. A new BT-Cond grid of Phoenix model atmospheres with updated opacities was presented by \citet{Allard_2012}, which spans $T_{\mathrm{eff}}=800-3000\,\mathrm{K}$. Figure \ref{fig:BT-COND_spec} shows comparisons of emission spectra from \texttt{ATMO} and BT-Cond for a selection of effective temperatures. Differences in the emission spectra can be seen in the $H$ and $K$ bands due to the updated $\mathrm{CH_4}$ line list and improved $\mathrm{H_2-H_2}$ collisionally induced absorption used by \texttt{ATMO}. Furthermore, for $\mathrm{T_{eff}=800\,K}$ differences can be seen in the $Y$ band at $\lambda\sim1\,\mu m$, likely due to different potassium broadening schemes (see Section \ref{sec:K_broadening}). 

\begin{figure}[t!]
\centering
\resizebox{1.0\hsize}{!}{\includegraphics{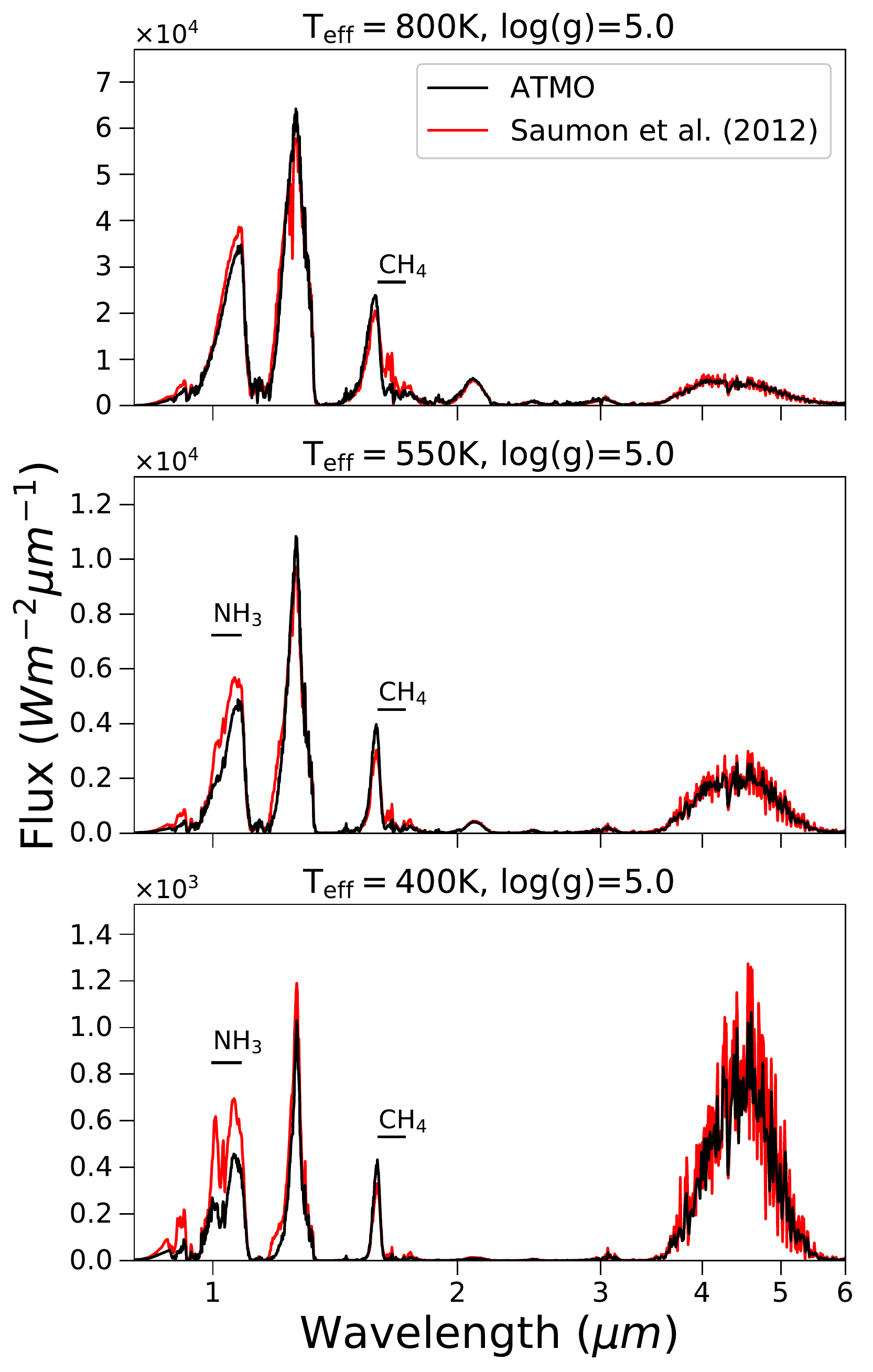}}
\caption{Synthetic infrared emission spectra from \texttt{ATMO} compared with models from \citet{Saumon_2012} for a range of effective temperatures. \label{fig:S12_spec}}
\end{figure}

\citet{Saumon_2012} (hereafter S12) also presented updated atmosphere models from the Saumon \& Marley group including improved $\mathrm{NH_3}$ and $\mathrm{H_2}$ opacities, for $T_{\mathrm{eff}}=300-1500\,\mathrm{K}$. We compare the infrared emission spectra predicted by \texttt{ATMO} against spectra from the S12 grid in Figure \ref{fig:S12_spec}, finding good overall agreement between the models, particularly in the $3.5-5.5\,\mu m$ flux window. Similarly to the comparisons with the BT-Cond grid, differences lie in the $\mathrm{CH_4}$ absorption band at $\lambda\sim1.6\,\mu m$ due to the updated line list used by \texttt{ATMO}. Further differences arise at lower $T_{\mathrm{eff}}$ within an $\mathrm{NH_3}$ absorption band at $\lambda\sim1\,\mu m$. Both the \texttt{ATMO} and S12 models use the same ExoMol line list from \citet{Yurchenko_2011}, meaning differences are likely due to differing $\mathrm{NH_3}$ abundances and condensation treatments in the models. We
note that further spectral differences between \texttt{ATMO} and the BT-Cond and S12 models in figures \ref{fig:BT-COND_spec} and \ref{fig:S12_spec} are likely due to discrepancies in the P-T profiles brought about by differing opacity sources impacting the temperature structure.

\subsection{Spectral comparisons with observations} \label{sec:obs_spectra}

\begin{figure}[h!]
\centering
\resizebox{1.0\hsize}{!}{\includegraphics{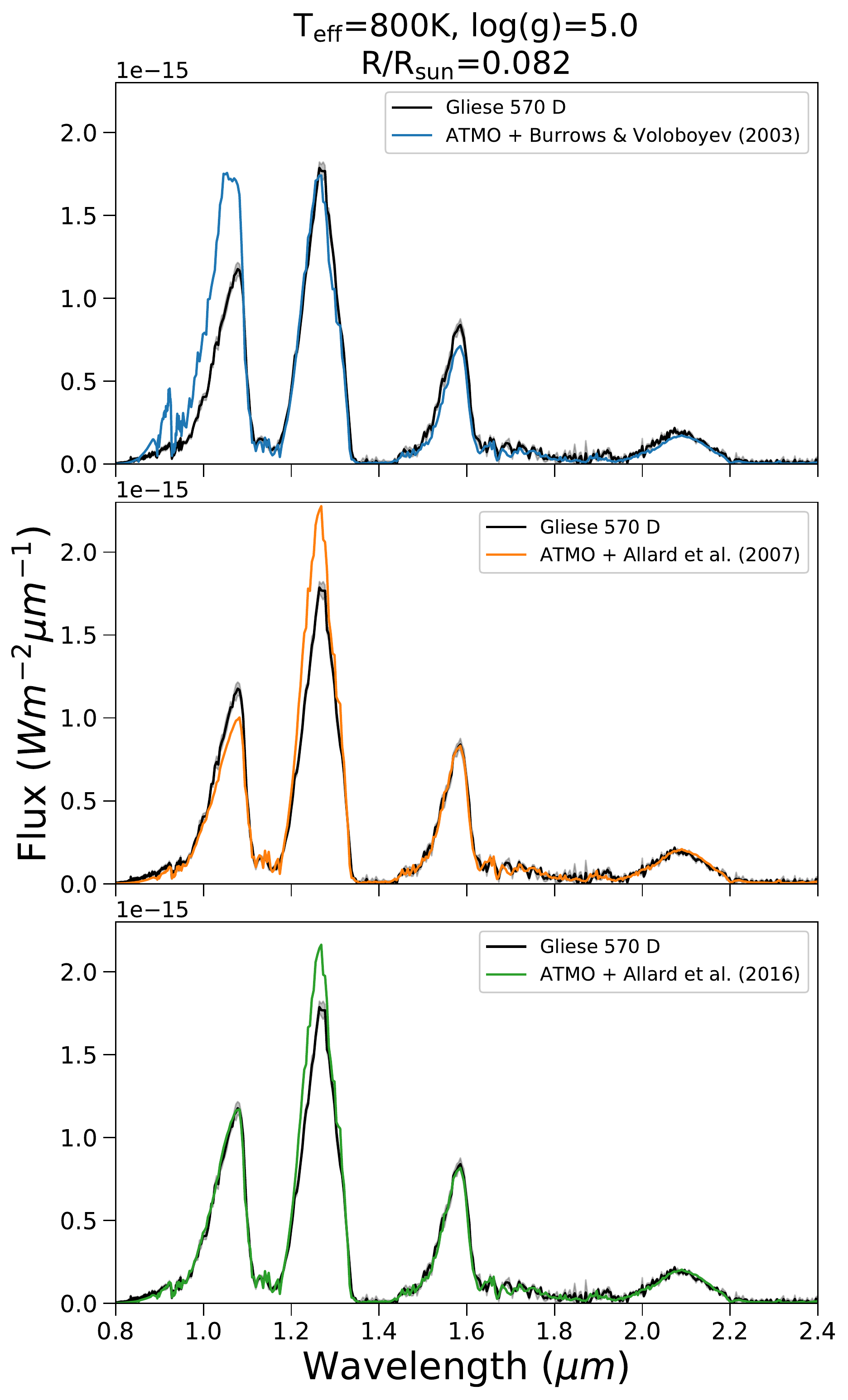}}
\caption{Model comparisons to the absolutely flux calibrated near-infrared spectrum of the T7.5 dwarf Gliese 570 D \citep{Burgasser_2000}. Three models are shown all calculated self-consistently using K resonance line broadening from BV03 (top), A07 (middle), and A16 (bottom), for $\mathrm{T_{eff}}=800\,\mathrm{K}$, $\log(g)=5.0$, and $\mathrm{R/R_{\odot}}=0.082$. \label{fig:Gliese_570_D}}
\end{figure}

\begin{figure*}[ht!]
\centering
\resizebox{\hsize}{!}{\includegraphics{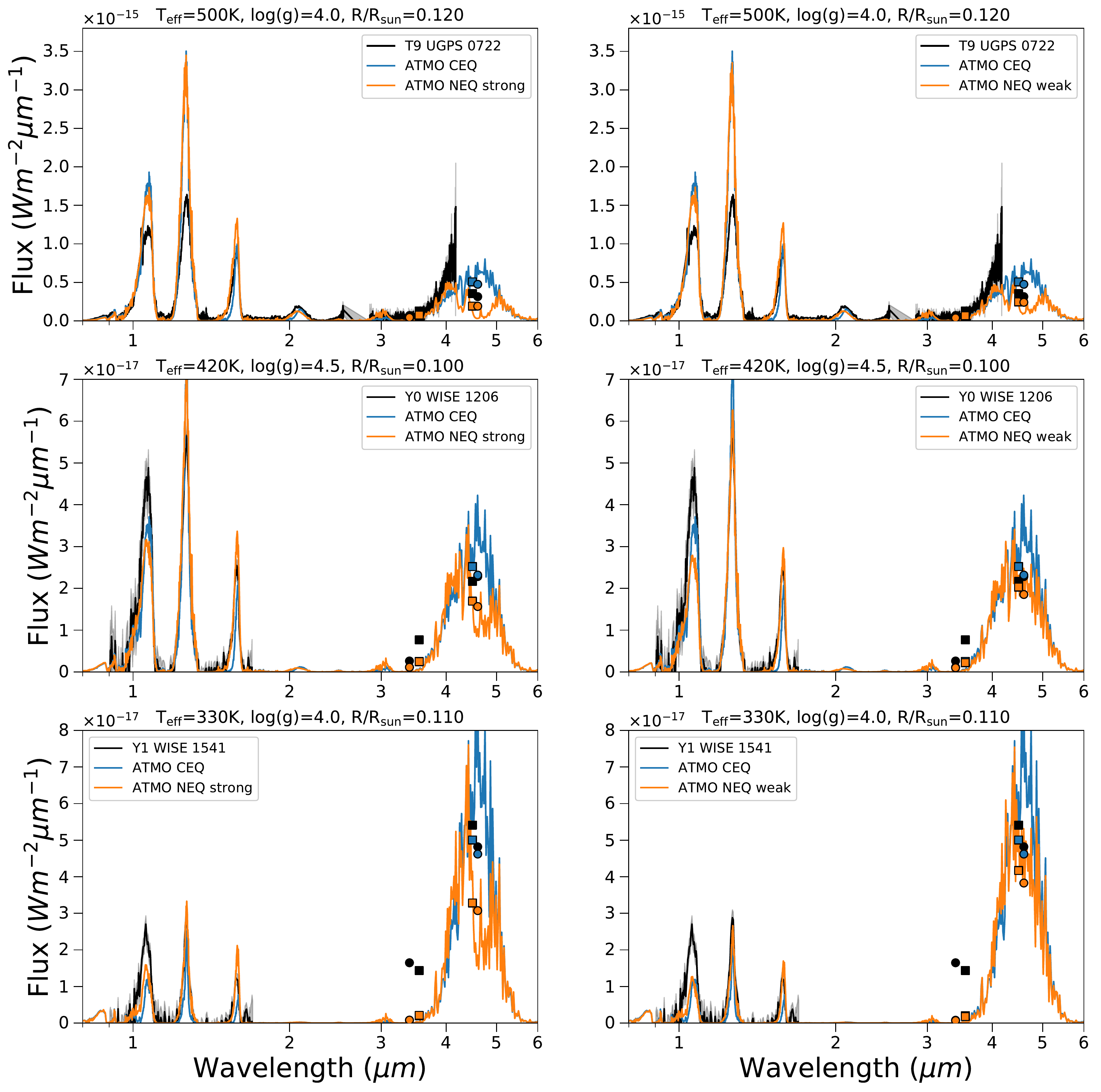}}
\caption{Comparison of the chemical equilibrium (blue) and non-equilibrium (orange) models of this work to sample spectra (black) forming a T--Y  spectral sequence. The left and right columns display non-equilibrium models with the strong and weak $K_{\mathrm{zz}}$ mixing relationships with surface gravity, respectively. \textit{Spitzer} IRAC photometry is plotted as squares and \textit{WISE} photometry as circles. The top panel shows the observed spectrum of the T9 dwarf UGPS 0722 \citep{Lucas_2010, Leggett_2012}, with \textit{Spitzer} IRAC and \textit{WISE} photometric points from \citet{Kirkpatrick_2012}. The middle and bottom panels show the $0.9-1.7\,\mu m$ \textit{HST} WFC3 spectra of the Y0- and Y1-type dwarfs WISE 1206 and WISE 1541 \citep{Schneider_2015}, with \textit{Spitzer} IRAC photometry also from \citet{Schneider_2015} and \textit{WISE} photometry from \citet{Cutri_2013}. \label{fig:spec_sequence}}
\end{figure*}

In Figures \ref{fig:Gliese_570_D} and \ref{fig:spec_sequence} we show comparisons of our models to spectra and photometry of cool T--Y-type  brown dwarfs. Our methodology here is to compare models to the data by eye, guided by values of $T_{\mathrm{eff}}$, $\log(g),$ and $R$ obtained from other studies in the literature and which are consistent with our new evolutionary tracks. Our by-eye comparison serves to illustrate model improvements and current shortcomings in reproducing cool brown dwarf spectra, and we leave more thorough grid fitting analyses to future work. 

Gliese 570 D is a late T dwarf companion to a ternary star system $\sim5.8\,\mathrm{pc}$ parsecs away from the sun \citep{Burgasser_2000, van_Leeuwen_2007}. It has a T7.5 spectral type and is one of the most thoroughly studied T dwarfs to date. Age indicators from the host star indicate an age in the range   $1-5\,\mathrm{Gyr}$ \citep{Geballe_2001, Liu_2007}. Gliese 570 D has been the target of a number of grid fitting studies, which have estimated $T_{\mathrm{eff}}=800-820\,\mathrm{K}$, $\log(g)=5.00-5.27$ and $L=2.88-2.98\,L_\odot$ \citep{Geballe_2001, Geballe_2009, Saumon_2006, Saumon_2012}. This object has also been used as a benchmark for brown dwarf retrieval studies, which obtain a slightly cooler $T_{\mathrm{eff}}=715\,\mathrm{K}$ and a surface gravity $\log(g)=4.8$ \citep{Line_2015, Line_2017}. Red-optical and near-infrared spectra are from \citet{Burgasser_2003, Burgasser_2004}.

We compare $T_{\mathrm{eff}}=800\,\mathrm{K}$, $\log(g)=5.0$ chemical equilibrium models calculated with different K resonance line broadening schemes to the red-optical and near-infrared spectra of Gliese 570 D \citep{Burgasser_2003, Burgasser_2004} in Figure \ref{fig:Gliese_570_D}. We find a radius of $\mathrm{R/R_{\odot}}=0.082$ provides the best match to the observed spectrum for this $T_{\mathrm{eff}}$ and $\log(g)$. Using our new evolutionary tracks, these parameters indicate an age of $5\,\mathrm{Gyr}$ and a mass of $46\,\mathrm{M_{Jup}}$ for Gliese 570 D, in agreement with previous works \citep{Saumon_2006}. We
note that non-equilibrium chemistry models do not impact the near-infrared spectrum within this wavelength range since the $K_{\mathrm{zz}}$ value is low for this high gravity.

The models with A16 K resonance line broadening provide the best match to the data. There is an excellent agreement in the $Y$ band where the redwing of the K resonance doublet influences the spectrum. In models with A07 K broadening the opacity in the redwing is too strong giving too little flux in the $Y$ band, whereas in models with BV03 broadening too much flux emerges in the $Y$ band due to the lower opacity in the K redwing. A further improvement in the models can be seen in the $H$ band, where the improved methane line list provides a much more satisfactory comparison to the data than models with a less complete  line list \citep{Saumon_2012}. The $K$ band is nicely reproduced due to the collisionally induced absorption from \citet{Richard_2012}, as previously shown in \citet{Saumon_2012}. The flux in the $J$ band is overpredicted by the model with A16 K broadening. We speculate that a more in-depth fitting study investigating the effects of metallicity and/or thermo-chemical instabilities may help further improve the fit in the $J$ band.

In Figure \ref{fig:spec_sequence} we show a comparison of our models calculated with equilibrium and non-equilibrium chemistry to spectra and photometry of objects spanning the T--Y  transition. We compare them to the T9 spectral standard UGPS 0722 \citep{Lucas_2010, Leggett_2012}, a well-studied cool dwarf that has been estimated to have $T_{\mathrm{eff}}=505\pm10\,\mathrm{K}$, a mass of $\mathrm{3-11\,M_{Jup}}$, and an age range between $60\,\mathrm{Myr}$ and $1\,\mathrm{Gyr}$ using the SM08 models \citep{Leggett_2012}. We compare $T_{\mathrm{eff}}=500\,\mathrm{K}$, $\log(g)=4.0$ chemical equilibrium and non-equilibrium models to this object, finding that these models  overpredict the flux in the $Y$ and $J$ bands at $\sim1.0\,\mu m$ and $\sim1.2\,\mu m$, respectively. This has been noted by other authors (e.g. \citet{Leggett_2012}), with sulfide clouds \citep{Morley_2012} or a reduced temperature gradient \citep{Tremblin_2015} invoked to redden the spectrum at these short near-infrared wavelengths. At longer wavelengths, the shape of the $K$ band at $\sim2.1\,\mu m$ appears to be better reproduced by the model including non-equilibrium chemistry. The \textit{Spitzer} IRAC channel 2 and \textit{WISE} $W2$ photometric points at $\sim4.5\,\mu m$ and $\sim4.6\,\mu m$, respectively, are lower than that predicted with the chemical equilibrium model, implying the presence of enhanced CO absorption brought about through vertical mixing in the atmosphere (see Figure \ref{fig:NEQ_spectra} and Section \ref{sec:NEQ_spec}). Both the strong and weak mixing non-equilibrium models overpredict the CO absorption in the IRAC ch2 and $W2$ bands, implying that the strength of vertical mixing is overestimated in our current model set-up. Decreasing the eddy diffusion coefficient $K_{\mathrm{zz}}$ further may improve the comparison to the photometric points in the $4-5\,\mu m$ flux window for this object.

Observations of the Y0- and Y1-type objects WISE 1206 and WISE 1541 \citep{Cushing_2011, Schneider_2015} are shown in the middle and bottom panels of Figure \ref{fig:spec_sequence}, respectively. Using the cloud-less models of SM08, \citet{Schneider_2015} estimate $T_{\mathrm{eff}}\sim400-450\,\mathrm{K}$ and $\log(g)=4.0-4.5$ for the Y0 object WISE 1206. \citet{Zaleksy_2019} ran retrieval analysis on a sample of Y dwarfs including WISE 1541, retrieving $T_{\mathrm{eff}}\sim325\,\mathrm{K}$ $\log(g)\sim5.0$ for this object, in line with comparisons to cloud-free forward models presented in \citet{Leggett_2013}. Here we compare $T_{\mathrm{eff}}=420\,\mathrm{K}$, $\log(g)=4.5$ models to WISE 1206 and $T_{\mathrm{eff}}=330\,\mathrm{K}$, $\log(g)=4.0$ models to WISE 1541. We use a lower value of the surface gravity for WISE 1541 than obtained by previous studies since $\log(g)=5.0$ does not agree with our evolutionary tracks at this $T_{\mathrm{eff}}$.

The $J$- and $H$-band brightness and shape is better reproduced by the non-equilibrium models for both objects. This is due to the quenching of $\mathrm{NH_3}$ reducing the opacity in these bands (see Figure \ref{fig:NEQ_spectra} and \citet{Tremblin_2015}). The strong mixing non-equilibrium chemistry model overpredicts the CO absorption in the $4-5\,\mu m$ flux window for the warmer WISE 1206 object, while the weaker mixing model better reproduces the photometric points in this wavelength range.  The strong and weak mixing non-equilibrium chemistry models both overpredict the CO absorption in the cooler WISE 1541 object, with the equilibrium model better reproducing the \textit{WISE} and \textit{Spitzer} photometry. 

The $Y$-band flux at $\sim1\,\mu m$ is underpredicted for WISE 1206 and WISE 1541 by the current models, an issue also seen by the model comparisons in \citet{Schneider_2015}. We note that the $\mathrm{K-H_2}$ opacity is important in this wavelength region; however, given the improvements to the K resonant line broadening outlined in Section \ref{sec:K_broadening}, we do not attribute this discrepancy to shortcomings in the K opacity. Instead, we note that reducing the K abundance by approximately an order of magnitude rectifies the difference between model and data in the $Y$ band. We therefore posit that the current modelling of the potassium chemistry, including potentially its condensation into KCl and/or the thermochemical data used to calculate the equilibrium abundances, is slightly incorrect. This should be investigated more thoroughly in future work.

\section{Discussion and conclusions} \label{sec:discussion}

We have presented our new \texttt{ATMO} 2020 set of substellar atmosphere and evolutionary models applicable to cool brown dwarfs and directly imaged giant exoplanets. Our atmosphere model grid is generated with our 1D code \texttt{ATMO}, and spans $T_{\mathrm{eff}}=200-3000\,\mathrm{K}$, $\log(g)=2.5-5.5$ with both equilibrium and non-equilibrium chemistry due to different strengths of vertical mixing. This grid of atmosphere models has been used as the surface boundary condition for the interior structure model to calculate the evolution of $0.001-0.075\,M_{\mathrm{\odot}}$ objects. We have highlighted numerous theoretical modelling improvements through comparisons to other model sets in the literature and comparisons to observational datasets in Section \ref{sec:Results}. We now discuss and conclude our work.


\subsection{Evolutionary tracks}

Our key result is that there are notable changes to the cooling tracks of substellar objects over previous works by B03 and SM08 (Figure \ref{fig:evol}). These changes are brought about by two major modelling improvements. First, the use of a new hydrogen and helium EOS \citep{Chabrier_2019} in the interior structure model has raised the hydrogen and deuterium minimum masses, causing changes in the cooling tracks around these masses (Figure \ref{fig:eos}). This new EOS includes quantum molecular dynamics calculations and predicts cooler, denser, and hence more degenerate objects than the previous semi-analytic EOS of \citet{Saumon_1995} used in previous works (Figure \ref{fig:interior}). We are planing to work on further improvements to the EOS, in particular improvements concerning the interaction between H and He, which could further alter the cooling curves.

The second major improvement concerns the line lists used for important molecular opacity sources such as $\mathrm{CH_4}$ and $\mathrm{NH_3}$ used in the model atmospheres. These line lists are taken primarily from the ExoMol group \citep{Tennyson_2018}, and include significantly more transitions required to accurately model the high-temperature atmospheres of brown dwarfs and giant exoplanets. These more complete line lists have added opacity to the 1D model atmosphere, changing the predicted emission spectra (Figures \ref{fig:BT-COND_spec} and \ref{fig:S12_spec}) and leading to warmer temperature structures (Figure \ref{fig:PT_profiles}). These warmer temperature structures are used as surface boundary conditions for the interior structure evolution model, and have changed the shape of the cooling tracks of substellar objects in the low-mass brown dwarf regime (Figure \ref{fig:evol}). For example, a $1\,\mathrm{M_{Jup}}$ object is cooler and less luminous at ages $\mathrm{<0.1\,Gyr}$, and warmer and brighter for ages $\mathrm{>0.1\,Gyr}$ compared to the previous AMES-Cond models. We note that line lists of important molecular absorbers are being continuously updated and improved, and we are working on integrating new $\mathrm{H_2O}$ \citep{Polyansky_2018} and $\mathrm{TiO}$ \citep{McKemmish_2019} line lists into our 1D atmosphere model. Investigating the impact of this updated opacity on the temperature structures, synthetic spectra, and cooling tracks will be the subject of future work.

\subsection{Potassium broadening}

Along with improvements to the molecular opacities, we have highlighted the improvement and importance of the pressure broadened potassium resonance doublet in Section \ref{sec:K_broadening}. We have implemented new K line shapes of A16 in our 1D atmosphere model \texttt{ATMO}. These new line shapes improve upon the line shapes of A07, including better determinations of the intermediate- and long-range part of the $\mathrm{K-H_2}$ interaction potential and spin-orbit coupling. We compare synthetic emission spectra calculated with these new K line shapes to others commonly used in the literature (BV03; A07), with our key finding being that there is a large impact and uncertainty on the predicted Y- and J-band flux due to these different line shapes. The large differences in opacity in the K far-red wing alters the flux through the Y band at $\sim1\,\mu m$, and leads to a redistribution of flux to longer wavelengths due to differences in the P-T profile, primarily impacting the J band. This redistribution of flux is only captured when  reconverging the P-T profile to find radiative-convective equilibrium when switching between opacity sources. 

To validate the new K line shapes of A16, we compare spectra computed with different K broadening line shapes to a typical late T dwarf Gliese 570 D (Figure \ref{fig:Gliese_570_D}). We find that the Y band at $\sim1\,\mu m$ is best reproduced by models including the new line shapes, with models computed with A07 and BV03 line shapes predicting too strong and too little absorption, respectively, in the far-red wing. Further work must be undertaken to validate these line shapes not only in the near-infrared, but also at red-optical wavelengths where these line shape calculations differ in the blue wing of the doublet (see Figures \ref{fig:K_opacity} and \ref{fig:K_spec}). The blue wing displays a satellite feature brought about by $\mathrm{K-H_2}$ quasi-molecular absorption that has previously been detected in the T dwarf $\epsilon$ Indi Ba \citep{AllardF_2007}, and can be a useful diagnostic of temperature and metallicity (A07). Accurately modelling the optical spectrum requires taking into account the pressure broadened line shapes of other alkali metals such as Na and Li. Recently, \citet{Allard_2019} presented improvements on the line shapes of the Na resonance doublet, finding a change in the blue wing of the doublet in the predicted emission spectra of self-luminous atmospheres. We are working to include these new Na resonance line shapes in our 1D atmosphere code, and future work to validate these improvements on the alkali opacity at red optical wavelengths must be undertaken. 

\subsection{Non-equilibrium chemistry}

In our new model set we have considered two chemistry scenarios, thermodynamic equilibrium and non-equilibrium due to vertical mixing. To model non-equilibrium chemistry we have consistently coupled the relaxation scheme of \citet{Tsai_2018} to our 1D atmosphere code, considering the non-equilibrium abundances of  $\mathrm{H_2O}$, CO, $\mathrm{CO_2}$, $\mathrm{CH_4}$, $\mathrm{N_2}$, and $\mathrm{NH_3}$. We adopt this relaxation scheme over a full kinetics network for computational efficiency and consistent convergence throughout the grid when solving for a self-consistent P-T profile. This relaxation scheme is more computationally efficient as it avoids the need to solve the large, stiff system of ordinary differential equations needed when using full chemical kinetics networks.

While we have considered the non-equilibrium abundances of the primary carbon- and nitrogen-bearing molecules, future models should include additional species thought to be impacted by vertical mixing. Non-equilibrium signatures of $\mathrm{HCN}$ may become apparent in high-gravity objects with vigorous mixing \citep{Zahnle_Marley_2014}, and $\mathrm{PH_3}$ and $\mathrm{GeH_4}$, both of which are signatures of vertical mixing in Jupiter's atmosphere, as well as $\mathrm{C_2H_2}$ and $\mathrm{CH_3D,}$ could impact the mid-infrared spectra of the coolest brown dwarfs \citep{Morley_2018}. Furthermore, present chemical kinetics models do not consider condensate species. As such, the models essentially assume that mixing of species into the upper atmosphere happens on timescales much shorter than condensation timescales. Such an assumption is important for $\mathrm{H_2O}$ and $\mathrm{NH_3}$,  which condense in the upper atmospheres of cool brown dwarfs in chemical equilibrium. Incorporating condensation timescales would involve combining kinetic cloud formation models such as the Helling \& Woitke model \citep{Woitke_2003, Woitke_2004, Helling_2006, Helling_2008b} with a gas-phase chemical kinetics scheme. While this coupling is technically challenging and beyond the scope of this work, coupled gas-cloud kinetics models are required to correctly determine the abundances of $\mathrm{H_2O}$ and $\mathrm{NH_3}$, which are critical species governing the temperature structure and thermal emission from cool Y dwarf atmospheres.

We demonstrate the impact of nitrogen and carbon non-equilibrium chemistry due to vertical mixing on the predicted emission spectra of cool T--Y-type objects (Figure \ref{fig:NEQ_spectra}). The quenching of $\mathrm{CH_4}$ and $\mathrm{NH_3}$ brightens the H band, and similarly to \citet{Hubeny_2007} we find the increased abundances and absorption of CO and $\mathrm{CO_2}$ suppress the flux in the $3.5-5.5\,\mu m$ flux window. The comparisons to the observed spectra of cool brown dwarfs in Figure \ref{fig:spec_sequence} indicate the quenching of ammonia is key in reproducing near-infrared emission of Y dwarfs, in agreement with the results of \citet{Leggett_2015, Tremblin_2015}. Comparisons of our new models to the $H-W2$ colours of cool brown dwarfs (Figure \ref{fig:K19_cmd}) and to photometric observations across the T--Y sequence  (Figure \ref{fig:spec_sequence}) also support the presence of non-equilibrium CO and $\mathrm{CO_2}$ abundances in the $3.5-5.5\,\mu m$ flux window. These results indicate that this wavelength region could be a useful tool in constraining the eddy diffusion coefficient $K_{\mathrm{zz}}$, the parameter commonly used to model vertical mixing in 1D atmosphere codes. The value of $K_{\mathrm{zz}}$ in brown dwarf and exoplanet atmospheres is a long-standing theoretical problem, and several attempts have been made to estimate it from numerical models (e.g. \citet{Moses_2011, Parmentier_2013, Zhang_Showman_2018, Zhang_Showman_2018b}). Since the abundances of CO and $\mathrm{CO_2}$ depend strongly on the value of $K_{\mathrm{zz}}$, the $3.5-5.5\,\mu m$ flux window could be a useful observational diagnostic of $K_{\mathrm{zz}}$ in cool brown dwarf atmospheres.

While non-equilibrium chemistry can reproduce several observational features of cool brown dwarfs, the red colours of late T dwarfs cannot be reproduced by cloud-free equilibrium or non-equilibrium models, similarly to the red late L dwarfs and the L-T transition. This is demonstrated by our comparisons to J-H colours in colour-magnitude diagrams (Figure \ref{fig:CMD_J-H}), and our spectral comparisons to the late T dwarf UGPS 0722 in Figure \ref{fig:spec_sequence}. Developing models which can reproduce this observed reddening will be the subject of future work.

\subsection{Future work}

Over the past half decade, a new theory has been developed suggesting that chemical transitions such as $\mathrm{CO \to CH_4}$ and $\mathrm{N_2 \to NH_3}$ in brown dwarf atmospheres can be responsible for triggering convective instabilities. This can reduce the temperature gradient in the atmosphere reddening the emission spectrum. Reductions in the temperature gradient through the effective adiabatic index $\mathrm{\gamma_{eff}}$ have been shown to reproduce several observed features of brown dwarfs, including the L-T transition \citep{Tremblin_2016}, extremely red young low-gravity objects \citep{Tremblin_2017b}, and the red colours of cool late T dwarf objects \citep{Tremblin_2015}. 

To investigate the potential mechanism at play reducing the temperature gradient in brown dwarf atmospheres, \citet{Tremblin_2019} generalised convection and mixing length theory to include diabatic processes through thermal and compositional source terms, demonstrating that a number of convective systems in the Earth's atmosphere and oceans derive from the same instability criterion. In brown dwarf atmospheres, the thermal and compositional source terms are represented by radiative transfer and $\mathrm{CO \to CH_4}$ or $\mathrm{N_2 \to NH_3}$ chemistry, respectively, with the convective instability driven by opacity and/or mean molecular weight differences in the different chemical states. The idealised 2D hydrodynamic simulations of \citet{Tremblin_2019} reveal that by including such source terms the temperature gradient can indeed be reduced to that required to qualitatively reproduce brown dwarf observations. Motivated by this, we aim to expand on this initial grid of atmosphere models by reducing the temperature gradient through the effective adiabatic index $\mathrm{\gamma_{eff}}$. Our key goal is to further study how radiative convection may evolve and influence the brown dwarf cooling sequence. We look forward to the good-quality, wide spectral coverage observational datasets that future instrumentation such as the James Webb Space Telescope will provide to aid in such studies.

\subsection{\texttt{ATMO} 2020 publicly available models}

The models presented in this work are publicly available for download\footnote{ \url{http://opendata.erc-atmo.eu}}\footnote{\url{http://perso.ens-lyon.fr/isabelle.baraffe/ATMO2020/}}. This includes P-T profiles, chemical abundances, and top of the atmosphere emission spectra, as well as evolutionary tracks with absolute magnitudes in a number of common photometric filters. We encourage users to contact the author if additional models or photometric filters are required, and while these are currently solar metallicity models, we plan to generate non-solar metallicity models in future work.

It should be noted that this grid is focused on cool T--Y-type brown dwarfs and that the range of validity of our models is $T_{\mathrm{eff}}\simle2000\,\mathrm{K}$, due to the calculations missing some sources of opacity (e.g. some hydrides, condensates) that form at higher temperatures and/or the modified atmosphere thermal gradients suggested for L dwarfs \citep{Tremblin_2016, Tremblin_2017a}. Such issues will be addressed in future studies. Therefore for brown dwarfs with $T_{\mathrm{eff}}\simgr2000\,\mathrm{K}$ (i.e. essentially massive objects younger than 100\,Myr with $M>0.03\,M_\odot$ or objects younger than 10\,Myr with $M>0.015\,M_\odot$), we suggest  using either the \citet{Chabrier_2000} or \citet{Baraffe_2015} models. We
note that the latter include an updated treatment of atmospheric convection.

\begin{acknowledgements}
We thank the anonymous referee for constructive comments which improved the manuscript. MWP ackowledges support through a UKRI-STFC studentship. The research leading to these results is partly supported by the ERC grant 787361-COBOM and by the STFC Consolidated Grant ST/R000395/1. P.T acknowledges supports by the European Research Council under Grant Agreement ATMO 757858. EH acknowledges support from a Science and Technology Facilities Council Consolidated Grant (ST/R000395/1). Calculations for this paper were performed on the University of Exeter Supercomputer, Isca, which is part of the University of Exeter High-Performance Computing (HPC) facility, and on the DiRAC Data Intensive service at Leicester, operated by the University of Leicester IT Services, which forms part of the STFC DiRAC HPC Facility (www.dirac.ac.uk). The equipment was funded by BEIS capital funding via STFC capital grants ST/K000373/1 and ST/R002363/1 and STFC DiRAC Operations grant ST/R001014/1. DiRAC is part of the National e-Infrastructure.
\end{acknowledgements}

\bibliographystyle{aa}
\bibliography{37381corr}

\end{document}